\documentstyle[11pt,epsfig,epsf]{article} 
\def\baselinestretch{1.3}
\newcommand{\ba}{\begin{array}}
\newcommand{\ea}{\end{array}}
\newcommand{\bd}{\begin{displaymath}}
\newcommand{\ed}{\end{displaymath}}
\newcommand{\be}{\begin{equation}}
\newcommand{\ee}{\end{equation}}
\newcommand{\bea}{\begin{eqnarray}}
\newcommand{\eea}{\end{eqnarray}}

\def\q2 {q^2}

\def\neu {\chi_1^0}

\begin{document}
\begin{flushright}
{\large MRI-P-000802\\ 
hep-ph/0009112}\\
\end{flushright}

\begin{center}
{\Large\bf  New Higgs signals from vector boson fusion in 
R-parity violating supersymmetry}\\[20mm]
Anindya Datta \footnote{E-mail: anindya@mri.ernet.in}, 
Partha Konar \footnote{E-mail: konar@mri.ernet.in} and  
Biswarup Mukhopadhyaya \footnote{E-mail: biswarup@mri.ernet.in}\\
{\em Mehta 
Research Institute,\\
Chhatnag Road, Jhusi, Allahabad - 211 019, India}
\end{center}

\vskip 20pt
\begin{abstract}
  
  We investigate signals of the lightest neutral Higgs in an
  R-parity violating supersymmetric model through the vector boson
  fusion mechanism. Assuming that R-parity is violated through lepton
  number, and locating regions in the parameter space where decays of
  such a neutral scalar into a pair of lightest neutralinos can be
  significant, we proceed to calculate the event rates for final
  states arising from decays of the neutralinos through both
  $\lambda$-and $\lambda'$-type interactions. Regions of the parameter
  space where each of these types of interactions can lead to
  detectable events are identified. The possibilities where such signals
  can be faked by other superparticles (squarks, gluinos, charginos and
  neutralinos) are also investigated. It is found that over a sizable
  region, one can obtain distinguishable signals of an intermediate mass
  neutral scalar from a study of the suggested final states at the
 Large Hadron Collider.
\end{abstract}

\vskip 1 true cm

\setcounter{footnote}{0}

\def\baselinestretch{1.8}

\section{Introduction}

To a large extent, the notion of a supersymmetric nature owes itself
to questions concerning the stability of the electroweak symmetry
breaking sector. It is, therefore, of natural interest to look for
some signature of supersymmetry (SUSY) \cite{SUSY} in the phenomenology
of the (lightest) Higgs boson if and when it is discovered in collider
experiments. In particular, if most of the SUSY particle spectrum is
on the heavier side, then it is of considerable importance to study
properties of the lightest neutral scalar and find out whether they
correspond to a SUSY scenario, and if so, {\em what kind of a
framework} it is.

Ironically, SUSY (at any rate in most of its incarnations) also
dictates that the lightest neutral Higgs be within a mass range of
about 140 $GeV$ \cite{higgs_mass}. Since the lower two-thirds of this
mass range is practically ruled out by data from the Large Electron
Positron (LEP) collider \cite{lep_higgs_bound}, we are left with the
so-called `intermediate mass range' in which a SUSY Higgs should
decidedly lie. The conventional method of producing such a neutral
scalar via gluon fusion seems to suffer from an abundance of
backgrounds for the dominant channels, and one has to depend only on
decays like $H \longrightarrow \gamma \gamma$ or $H \longrightarrow V
V^{*}$ (where $V = W, Z$) to uncover its presence \cite{higgs_search}.
Under the circumstances, a detailed investigation of the couplings and
other properties of the Higgs becomes a rather difficult proposition.

A parallel channel explored for the discovery of a Higgs at a hadron
collider (such as the upcoming Large Hadron Collider (LHC) at CERN)
is the fusion of $W(Z)$ bosons emitted from quark pairs \cite{w-fuse}.
Though it was originally discussed in the context of a heavy Higgs
\cite{w-fuse_hev_higgs}, its usefulness in finding signals of an
intermediate mass Higgs \cite{int_mass_higgs}, too, has been
established.  Tagging of the energetic forward jets associated with
such a process, together with the absence of hadronic activity in the
large rapidity gap between them, \cite{forward_jet} can considerably
reduce backgrounds for final states ensuing from different Higgs decay
channels. The process has been studied, both in the context of the
standard model (SM) and the minimal SUSY standard model (MSSM), for
channels such as $\tau \overline{\tau}$, $b \overline{b}$, $\gamma
\gamma$ and for associated $HW$ production processes
\cite{zeppen_rain}, and the advantages compared to the gluon fusion
mechanism have been reiterated. The observation of these various final
states arising from the decay of neutral Higgs will undoubtedly go a
long way in establishing the  properties of the latter \cite{mlhc}.

In this paper, we focus our attention on R-parity violating SUSY
theories \cite{susy_rp}.  There is nothing that forbids the
multiplicative quantum number $R$, defined as $R = (-)^{3B + L + 2S}$,
from being violated in SUSY so long as {\it one} of baryon (B) or
lepton (L) number is conserved. On the other hand, the violation of
R-parity makes the lightest SUSY particle (LSP) unstable, thereby
altering many of the conventional signals of the MSSM
\cite{rp_signals}.  Here we argue that if R-parity is violated, the
small but non-negligible fraction of neutral scalars decaying into the
$\chi_1^0 \chi_1^0$ channel (where $\chi_1^0$ is the lightest
neutralino, the LSP in most cases), followed by decays of the
$\chi_1^0$ into three fermions, will lead to useful signals of the
Higgs via the vector boson fusion (VBF) mechanism.

This channel has been discussed earlier in the context of MSSM as an
invisible decay mode of the Higgs \cite{djouadi,madhu}. As almost all
of the parameter space that could make this the dominant decay mode
has been ruled out by LEP data, its relevance in the context of
R-conserving theories is perhaps not very high any more. However, we
want to show that even a branching ratio of a few per cents (or less)
for this channel can make it detectable if the VBF technique is
employed. This not only gives us the source of a substantial signal
for an intermediate mass Higgs, but also allows the measurement of the
Higgs coupling with a neutralino pair if R-parity is violated.

The production of the neutral scalar via gluon fusion can also give
rise to new signals if it decays into LSP pairs which in turn 
have three-body decays. However, in such cases it is very difficult to
distinguish the Higgs signals against the backdrop of numerous superparticle
cascades, most importantly those originating from squark or gluino
pair production, all leading to the production of LSP pairs. As we shall
show here, the VBF signals give us tags with which we can eliminate
the cascade `backgrounds' quite effectively.

We confine ourselves to R-parity violation in terms of lepton number only.
In section 2, we present an investigation of the SUSY parameter space and try
to outline the region where the two-neutralino decay mode for the lightest
Higgs can have any hope of detection. The signals corresponding to
the so-called $\lambda$-and $\lambda^{'}$-type couplings are discussed in
sections 3 and 4. We have also looked into the possibilities of
the Higgs signals being faked by strongly interacting superparticles (i.e.
squarks and gluinos) as well as 
charginos or neutralinos. After a detailed estimate of such `backgrounds',
we identify in these two sections the regions in the parameter space 
where the corresponding signals have a strong chance
of standing out. We summarise and conclude in section 5.

\section{Analysis of the parameter space}

Let us first take a close look at the parameter space of the theory and
try to identify the regions where the  $\chi_1^0   \chi_1^0$ decay mode 
for the lighter neutral Higgs can have a branching ratio of one per cent or 
more. In doing so, we recall that in a general R-parity violating SUSY
model, the following terms are added to the MSSM superpotential, written
in terms of the quark, lepton and Higgs superfields \cite{barger}:

\begin{equation}
W_{\not R} = \lambda_{ijk} {\hat L}_i {\hat L}_j {\hat E}_k^c +
\lambda_{ijk}' {\hat L}_i {\hat Q}_j {\hat D}_k^c +
\lambda_{ijk}''{\hat U}_i^c {\hat D}_j^c {\hat D}_k^c + \epsilon_i {\hat
L}_i {\hat H}_2
\end{equation}
where the $\lambda''$-term causes B-violation, and the remaining ones,
L-violation. In order to suppress proton decay, it is customary to
have one of the two types of nonconservation at a time. Here we will
consider only lepton number violating effects. Furthermore, we simplify
our analysis by keeping, in turn, only the $\lambda$-and
$\lambda'$-type interactions. Experimental limits on the individual
couplings can be found in the literature \cite{rp_limits}. The
presence of such interactions (which lead, among other things, to the
instability of $\chi_1^0$) affects the MSSM parameter space allowed by
the LEP data, mainly through chargino search limits
\cite{chargino_limit}. They, however, do not have any noticeable
effect on the Higgs potential, and therefore the results of analyses
corresponding to MSSM can be taken over directly for our purpose. It
may be worthwhile to note here that when the bilinear terms
$\epsilon_i {\hat L}_i {\hat H}_2$ are included \cite{roybabu}, the
presence of explicit Higgs-slepton mixing alters the character of the
potential.  Nonetheless, the conclusions reached by us are not
drastically altered even upon the inclusion of such terms. We shall
comment on this again in section 5.

We assume gaugino mass unification at a high energy scale, so that all
masses and mixing angles in the chargino-neutralino sector are fixed when 
we specify the SU(2) gaugino mass $M_2$, the Higgsino mass parameter $\mu$,
and $\tan \beta$, the ratio of the vacuum expectation values (vev) of the
two Higgs doublets. No
supergravity (SUGRA) framework has been  postulated, so that 
the squark and slepton masses (and also $\mu$) can be 
treated essentially as free parameters. 

In the Higgs sector, the physical states are comprised of two neutral
scalars ($h,~H$), a neutral pseudoscalar ($A$) and two mutually
conjugate charged scalars ($H^\pm$). At tree level, all these masses
and also the neutral scalar mixing angle ($\alpha$) get completely
determined once the pseudoscalar mass ($m_A$) and $\tan \beta$ are
specified.  In addition, they are influenced by the top quark mass,
the squark masses and the trilinear SUSY breaking parameter $A_t$ when
radiative corrections to the potential are taken into account. Here we
have used the full one-loop corrected Higgs potential
\cite{higgs_one_loop} to determine the various masses and the mixing
angle $\alpha$. Our results for different Higgs masses and
corresponding branching ratios  are consistent with those in
\cite{HDECAY}.

\begin{figure}[ht]
\centerline{
\epsfxsize=6.5cm\epsfysize=6.0cm
                     \epsfbox{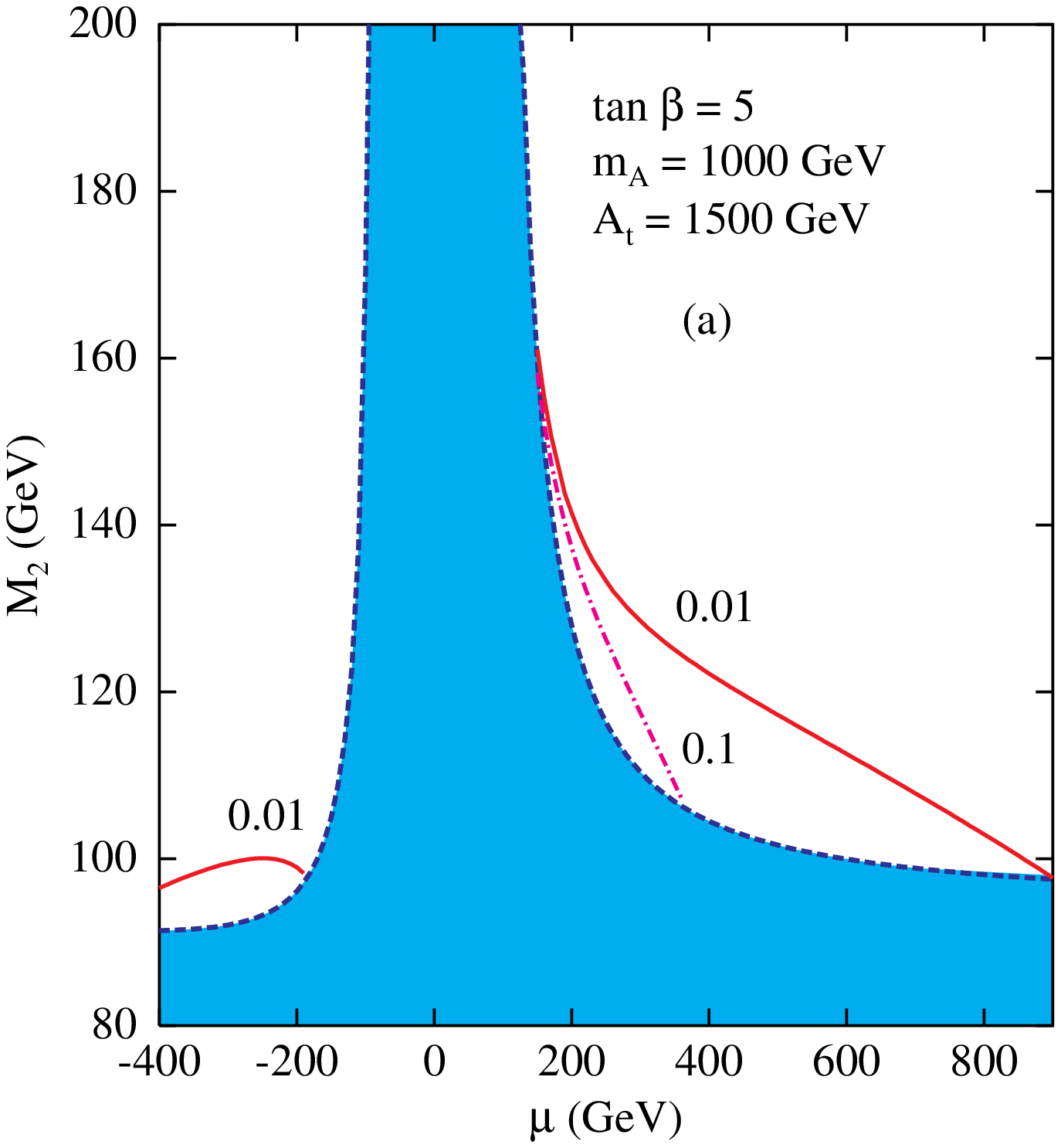}
\epsfxsize=6.5cm\epsfysize=6.0cm
                     \epsfbox{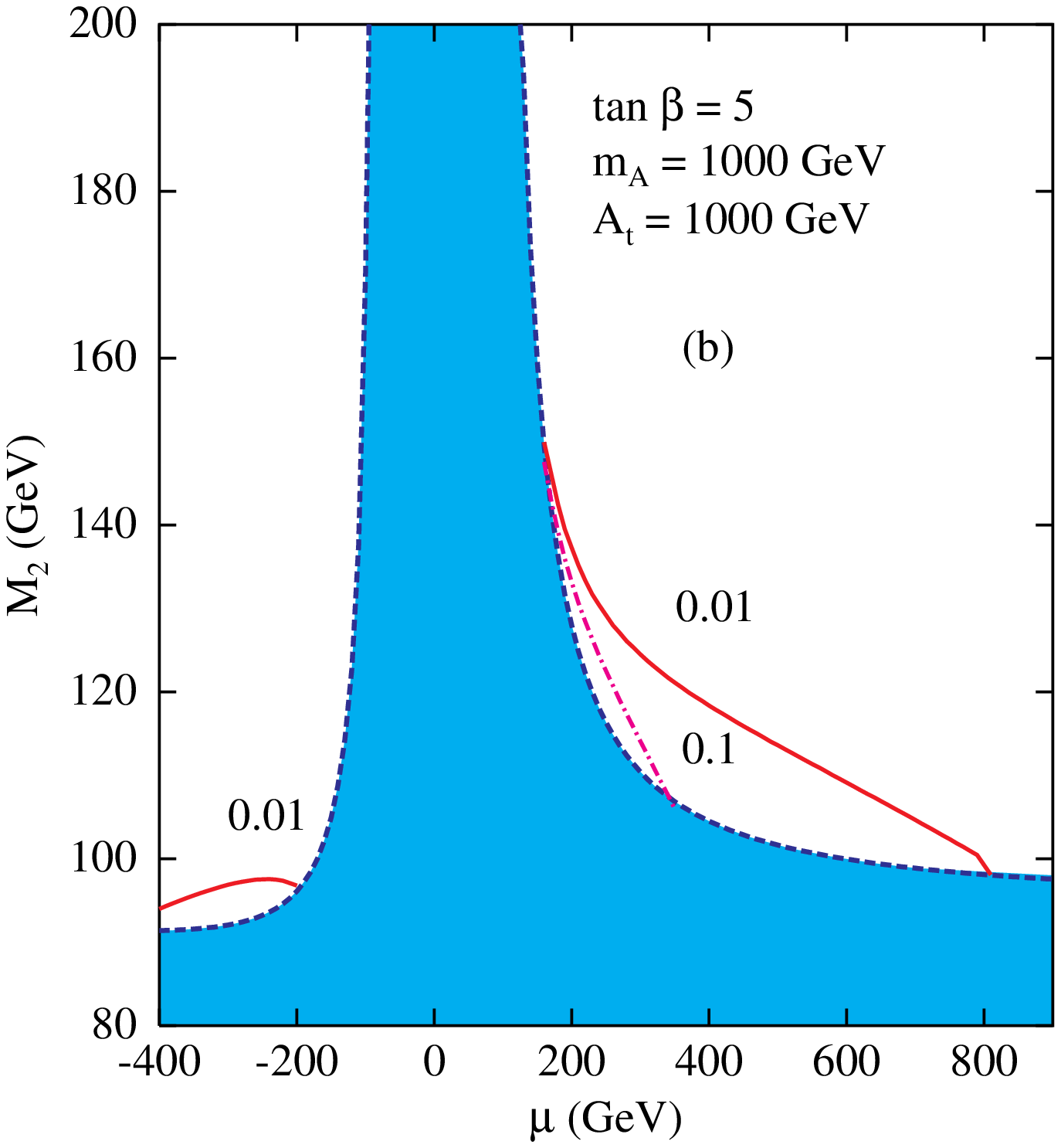}
}

\vspace*{2ex}
\centerline{
\epsfxsize=6.5cm\epsfysize=6.0cm
                     \epsfbox{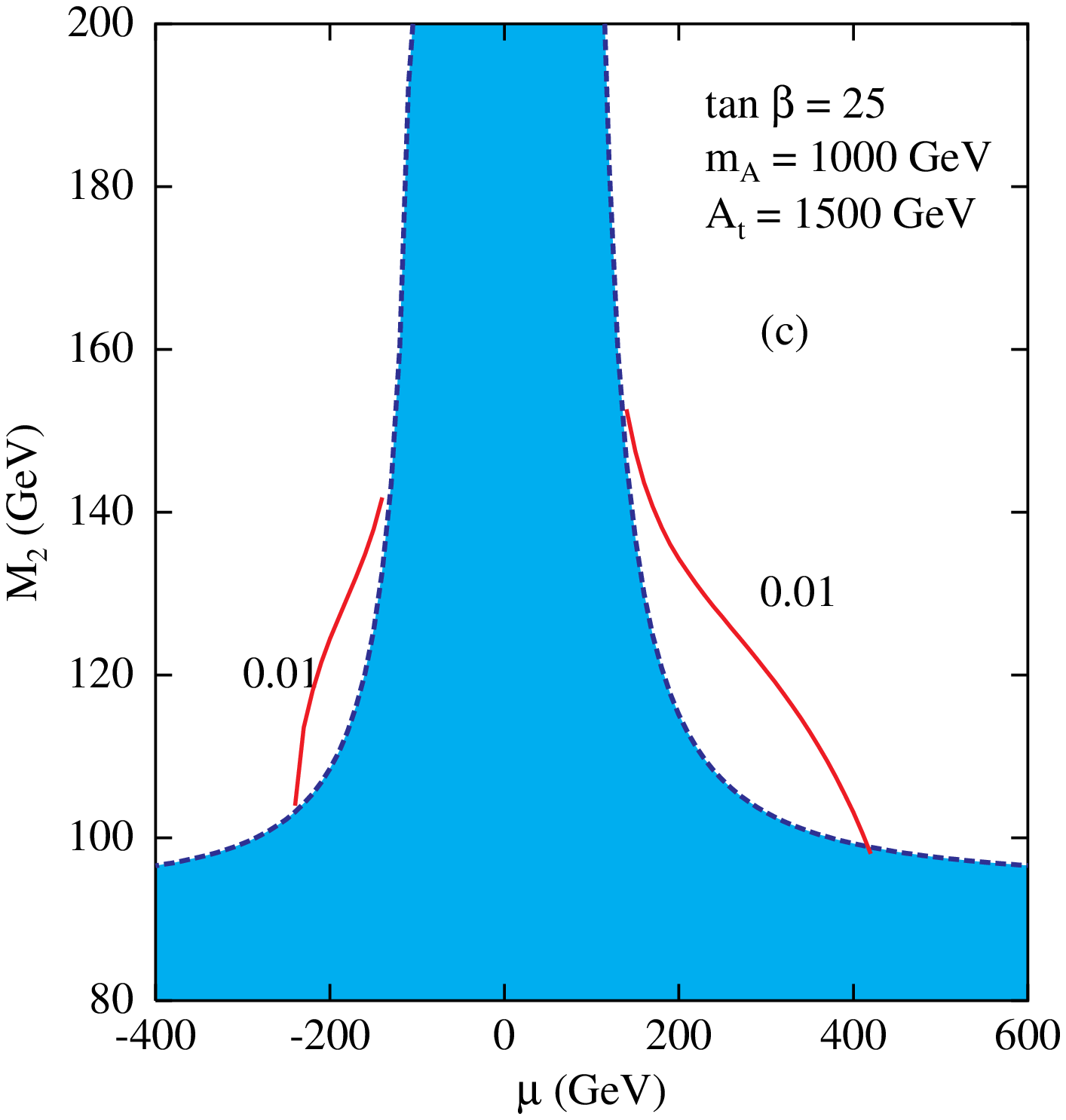}
\epsfxsize=6.5cm\epsfysize=6.0cm
                     \epsfbox{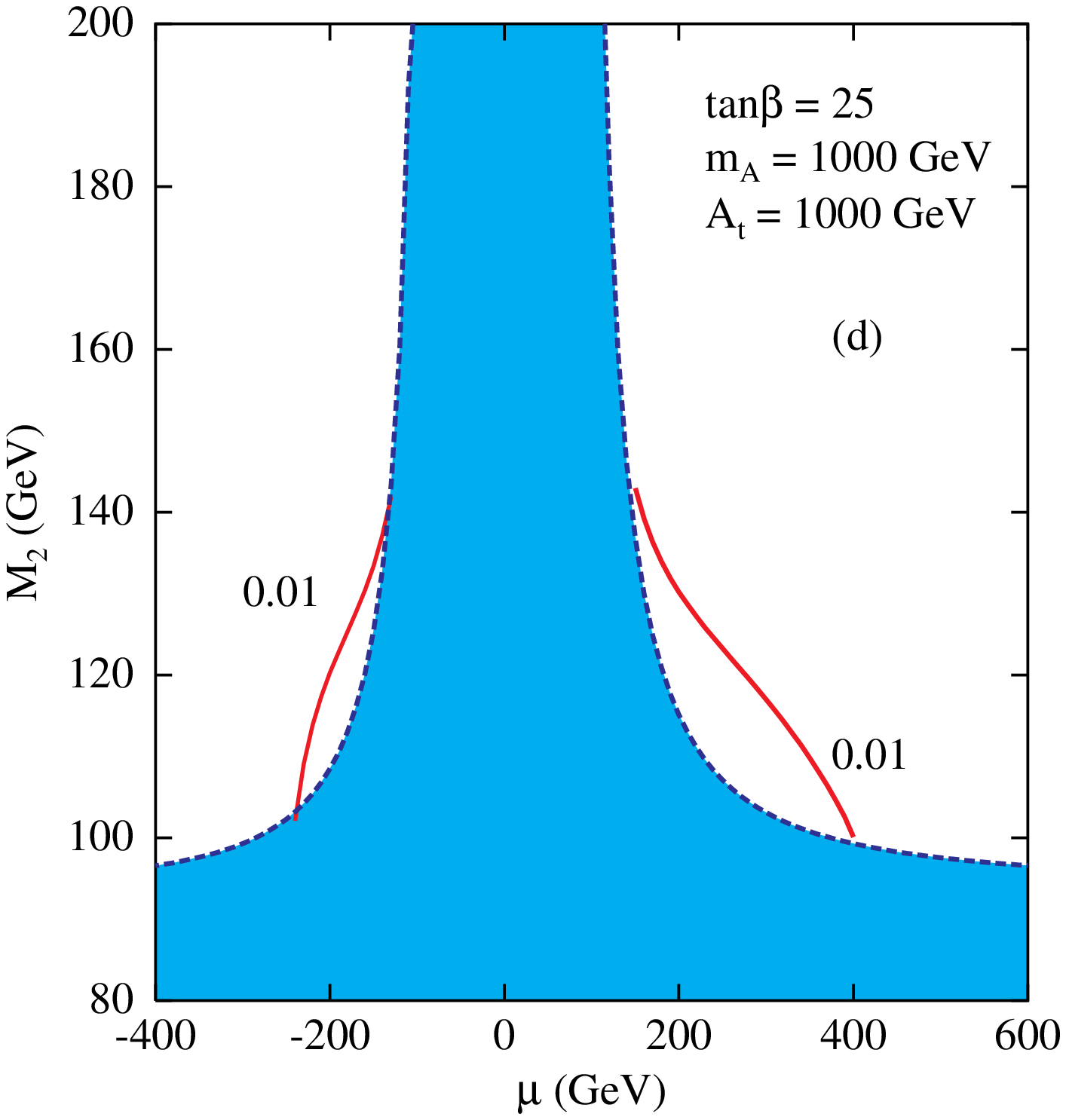}
}
\caption{ {\it Contours of constant branching ratio for  $ h \rightarrow 
    \neu \neu$ in the $\mu - M_2$ plane. The four panels are for 
    different choices of $A_t$ and $\tan \beta$.  }}
\label{fig:br_mu_m2}
\end{figure}

The decay width of the lightest neutral scalar $h$ into two lightest
neutralions is given by \cite{djouadi}.

\begin{equation}
\Gamma (h \longrightarrow \chi_1^0   \chi_1^0) = 
{\frac{G_F m_W^2 m_h}{2\sqrt{2} \pi}} |\Delta_{11}|^2 (1  - 
4 m_{\chi_1^0}^2/m_h^2)^3
\end{equation}

\noindent
where 

\begin{equation}
\Delta_{11} = (N_{12} - N_{11} \tan \theta_W)(N_{13} \sin \alpha  +
N_{14} \cos \alpha) 
\end{equation}

\noindent
N being the neutralino mixing matrix in the basis
($\tilde{B}, \tilde{W_3}, \tilde{H_1}, \tilde{H_2}$). As is evident from the
expression, the decay requires contributions from the gaugino components
of one neutralino and Higgsino components of the other. Thus the branching
ratio is expected to go down when either of $M_2$ and $\mu$ becomes
large compared to the other, so that either the gaugino or the Higgsino
components may fall appreciably.

In figures 1(a -d), we show contours of different branching ratios in the
$\mu$-$M_2$ plane. The region disallowed by LEP data (upto the 202 $GeV$ run)
has been shaded out in each graph. Clearly, even within the LEP-allowed 
regions, a branching ratio of the order of 1\% and ranging up to 10\% are
possible. There is a predictable decrease in the branching ratio as one moves
outwards in each case, since, in addition to the reason given in the previous
paragraph, the $\chi_1^0$ mass rises when both $M_2$ and $\mu$
are increased.

A comparison between 1(a) and 1(c) (as also between 1(b) and 1(d)) shows that 
larger values of $\tan \beta$ tend to suppress the branching ratio for
the $\chi_1^0 \chi_1^0$ channel, an effect resulting mainly from the 
enhancement of the $b \bar{b}$ coupling of $h$.  
\begin{figure}[ht]
\centerline{
\epsfxsize=6.5cm\epsfysize=6.0cm
                     \epsfbox{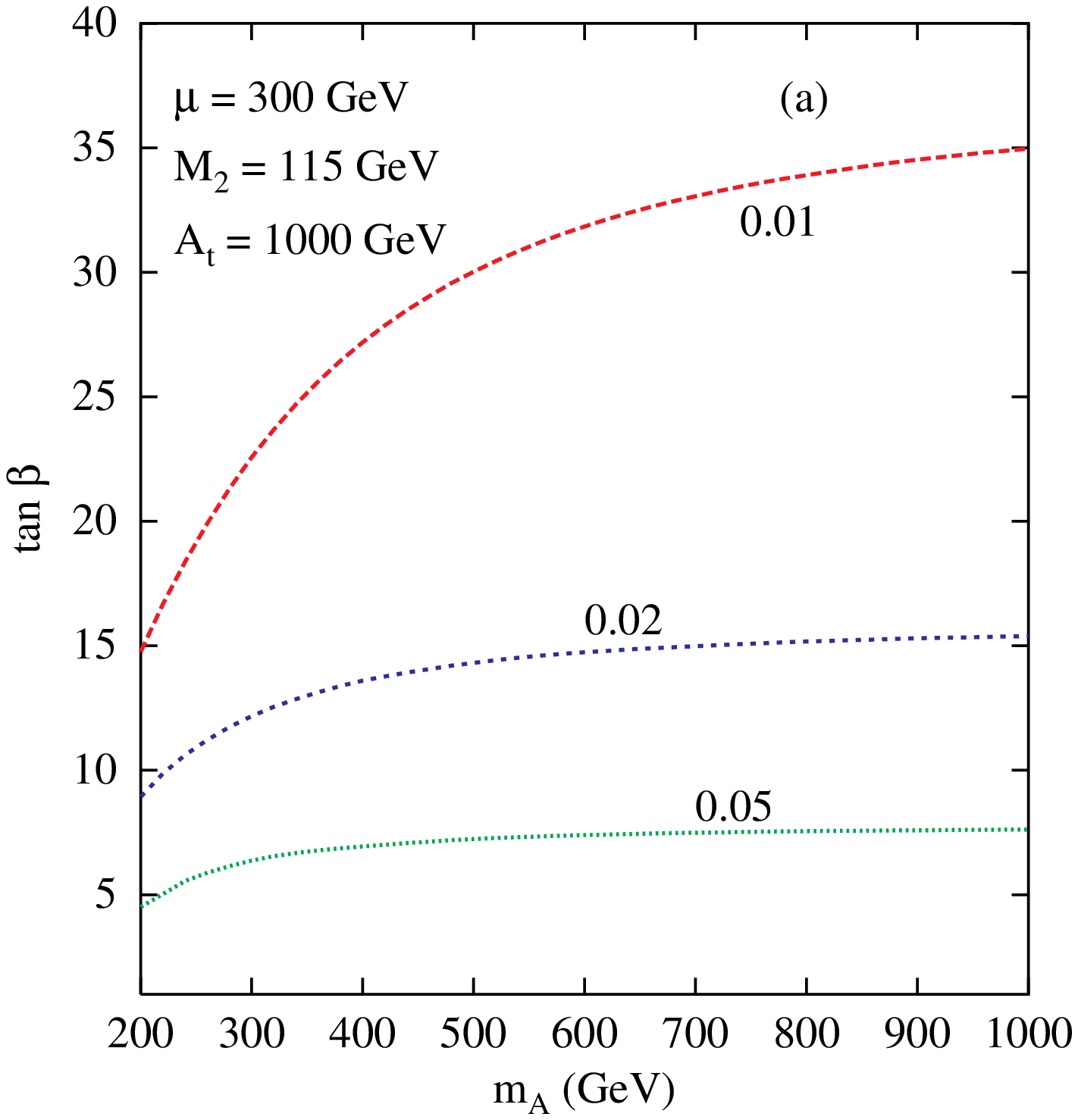}
        \hspace*{-1ex}
\epsfxsize=6.5cm\epsfysize=6.0cm
                     \epsfbox{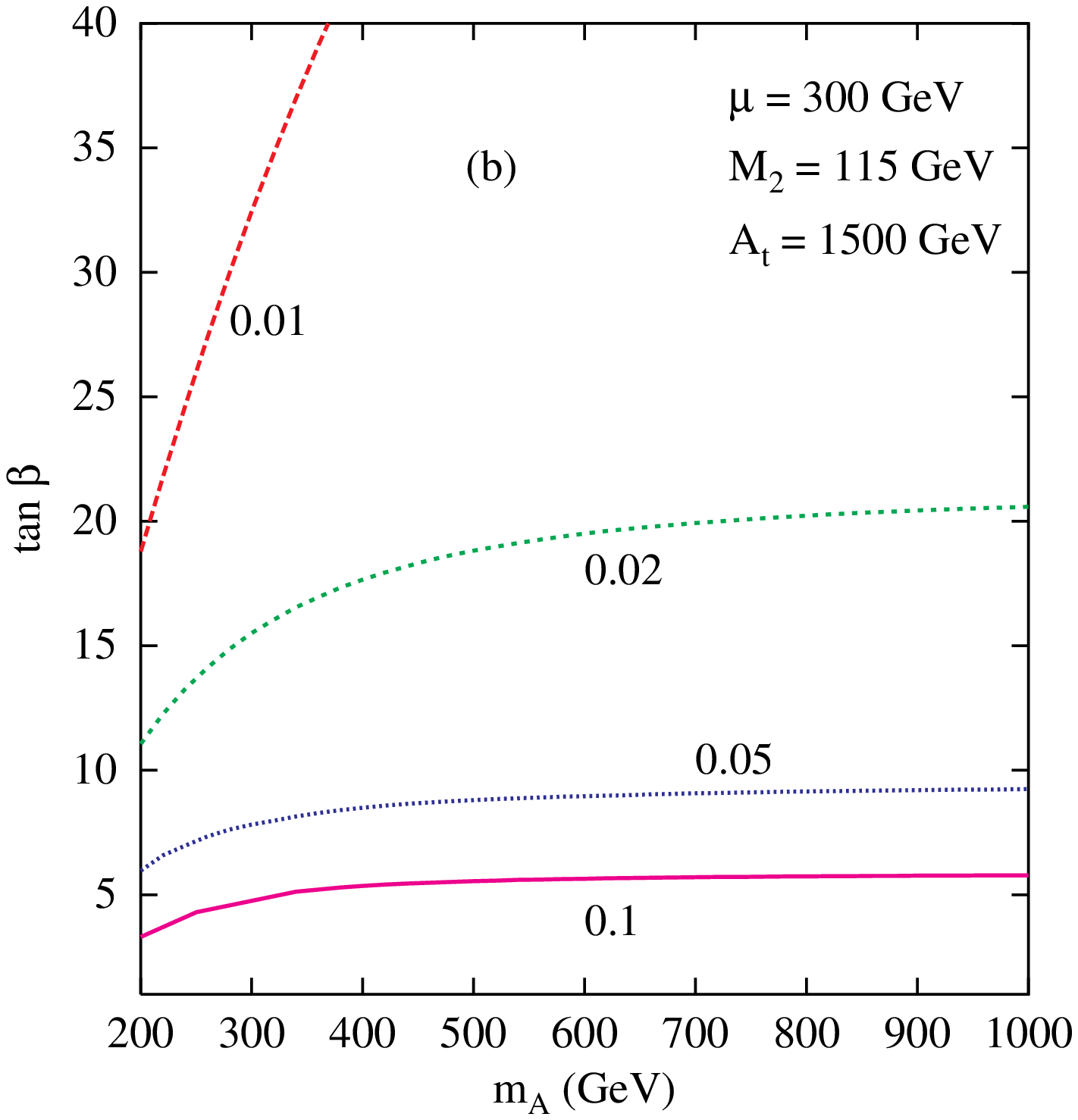}
}

\caption{{\it Contours of constant branching ratio for  $h \rightarrow
\neu \neu$ in the $m_A - \tan\beta$ plane.} 
}
\label{fig:br_ma_tanb}
\end{figure}

On the other hand, the decay of our interest is seen to be boosted if
one has a larger value of the trilinear SUSY breaking parameter $A_t$.
As has already been mentioned, the latter has a crucial influence on
the one-loop corrections to the scalar potential, thereby affecting
not only the neutral scalar mass but also the mixing angle $\alpha$.
However, $A_t$ is also constrained from considerations such as the
absence of flavor-changing neutral currents, and more stringently,
from the requirement to prevent charge and color breaking as well as
instability of the scalar potential \cite{ccb}.  Keeping all these
constraints in view, the value of $A_t$ can be as high as 1 TeV and
can even go up to 1.5 TeV, but with a simultaneous increase in the
sfermion masses. Thus higher branching ratios for $h \longrightarrow
\chi_1^0 \chi_1^0$, triggered by $A_t$, seem to be more likely when
the squarks and sleptons are close to 1 TeV.

In figures \ref{fig:br_ma_tanb}, we show the branching ratio contours in the
$\tan \beta - m_A$ plane. Together with $A_t$, these two variables fix 
the mass of the decaying $h$. The figures show that there is a marginal
increase in the branching ratio as one increases  $m_A$. This effect is more
pronounced for higher values of $\tan \beta$. The reason behind this
is the fact that in the region of our interest, the decay width in    
equation (2) is controlled by the terms proportional to $\cos \alpha$ which,
as can be easily checked, increases slowly with $m_A$. However, the same terms
are also proportional to the quantity $N_{14}$ which is larger for
large $\tan \beta$, thereby leading to the features observed in the
figures.    

The above analysis thus leads us to the conclusion that over a 
sizable region of the parameter space, the two-neutralino decay mode
of an intermediate mass Higgs can have a branching ratio ranging from
1 to 10 per cent and can occasionally go up to 20 per cent as well. 
As we shall see in the following sections, such values
can yield detectable and background-free events at the LHC when the
lightest neutralino is unstable.

\section{Signals with $\lambda$-type interactions}

Since there are thirty-six independent $\lambda$-and $\lambda'$-type couplings 
which are {\it a priori} unrelated, a transparent analysis is possible
when only some of them are considered at a time. Here we assume the
presence of just {\it one}  $\lambda$-type interaction (say, $\lambda_{212}$)
which can lead to the decay $\chi_1^0 \longrightarrow l \bar{l'} \nu$ for
the lightest neutralino, with $l, l'~= e, \mu$.

The experimental limits on the interactions of the above kind can be found, 
for example, in reference \cite{rp_limits}. However, our predicted number of events will  
be independent on the actual value of the coupling so long as there is just
one coupling driving the decay of $\chi_1^0$ (and we impose identical event
selection criteria for both electrons and muons). 

Thus the type of events we are predicting here can be described as

\begin{equation}
qq \longrightarrow qqh \longrightarrow qq \chi_1^0 \chi_1^0 
\longrightarrow qq + 4l + \not{E_T}
\end{equation}

\noindent
the missing transverse energy coming from the neutrinos produced in
three-body decays of the neutralino. The absence of color exchange between the
quarks leads to a suppression of hadron production in the  central region, 
so that a `central jet 
veto' (whose efficacy in eliminating backgrounds can be established by 
looking into the VBF process along with one-parton emission) can be applied
for final states like the ones under consideration here. The two quarks
jets are highly energetic and in high-rapidity regions, with
a large rapidity gap in between, where the four leptons resulting from
Higgs decay are expected to lie. It is by tagging these forward jets
that one can trace the origin  of the neutralino decay products
to the neutral scalar $h$, thereby distinguishing them from conventional
superparticle cascades started by the production of squarks or
gluinos through strong interactions.

Our calculation is based on a parton level Monte Carlo for pp
collisions with $\sqrt{s}~=~14~TeV$. Here as well as in the next
section, we have set a degenerate squark mass of 300 $Gev$ and 
a degenerate slepton mass of 200 $Gev$. We have used the CTEQ4L
\cite{cteq} parton distribution functions. The lowest order tree-level
matrix elements for both $WW$ and $ZZ$ fusion processes have been
used. We have not included QCD corrections which are usually
rather modest \cite{qcd_corr}. The jet and lepton energies have been
further smeared using Gaussian functions, with half-widths ($\Delta
E$) given by \cite{smear}

\begin{equation}
\Delta E ~=~ 0.15 \sqrt{E} ~+~ 0.01E
\end{equation}

\noindent
for leptons, and

\begin{equation}
\Delta E ~=~ 0.4 \sqrt{E} ~+~ 0.02E
\end{equation}

\noindent
for jets.

\begin{figure}[ht]
\centerline{
\epsfxsize=6.5cm\epsfysize=6.0cm
                     \epsfbox{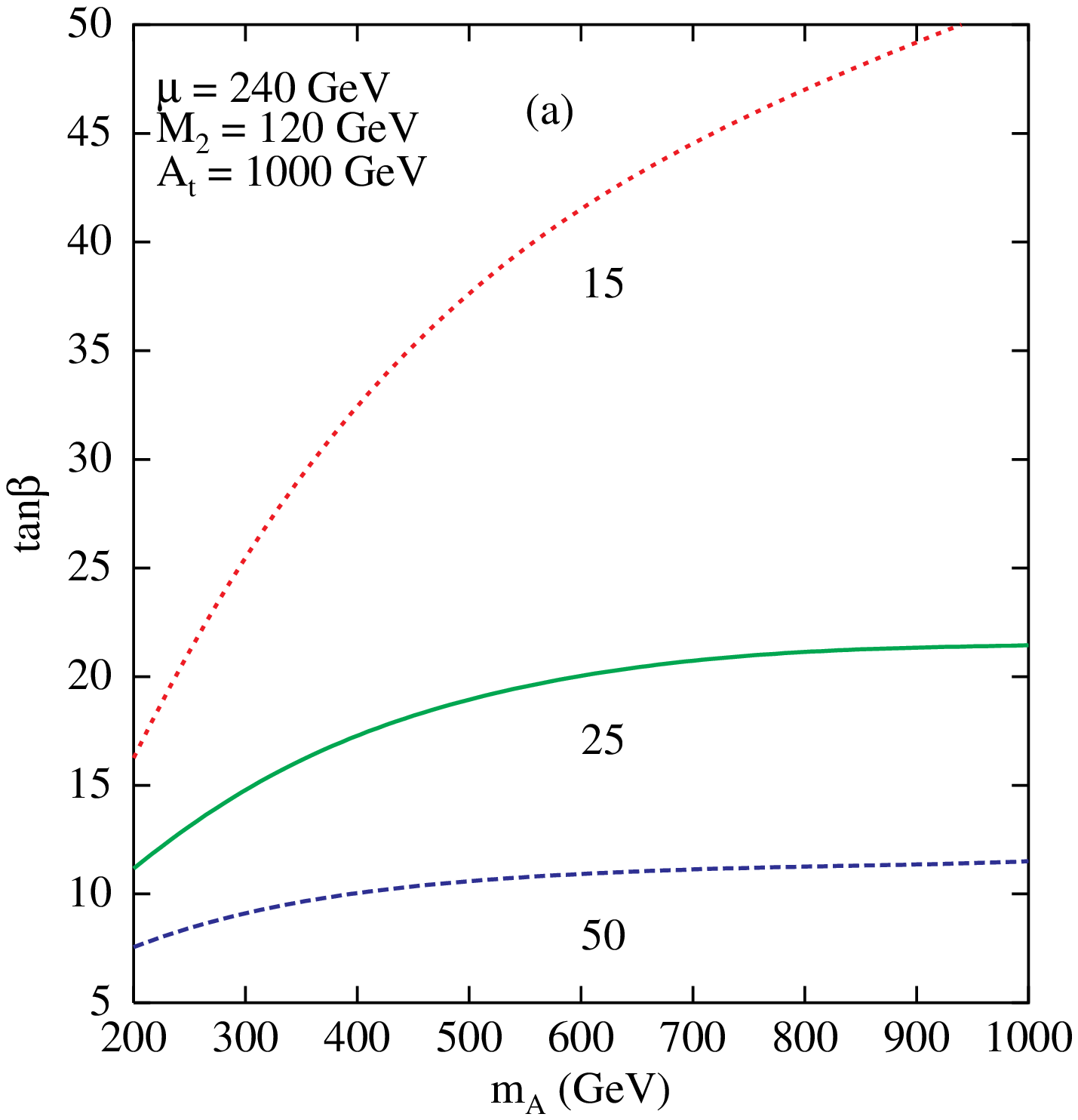}
\epsfxsize=6.5cm\epsfysize=6.0cm
                     \epsfbox{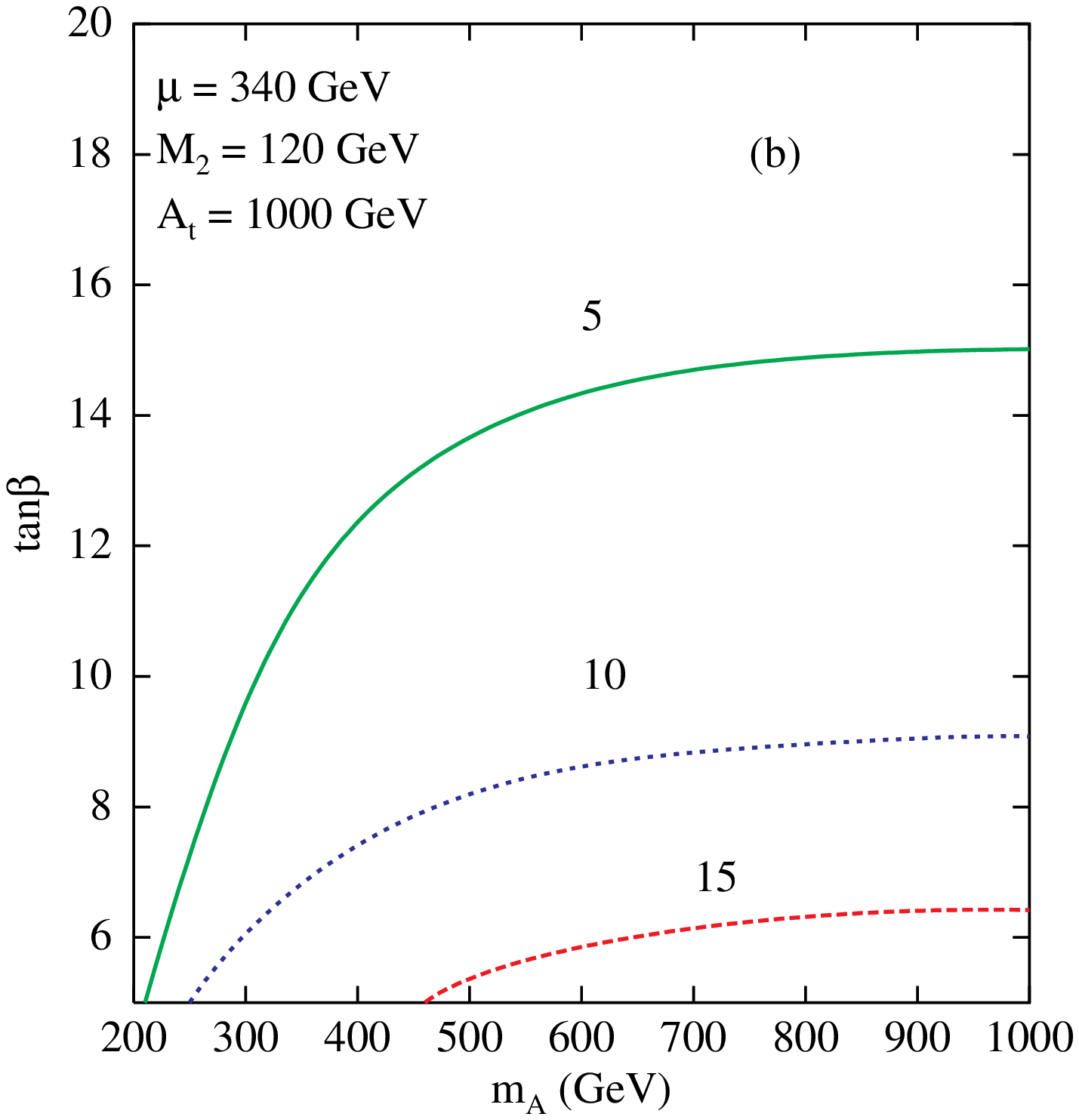}
}

\centerline{
\epsfxsize=6.5cm\epsfysize=6.0cm
                     \epsfbox{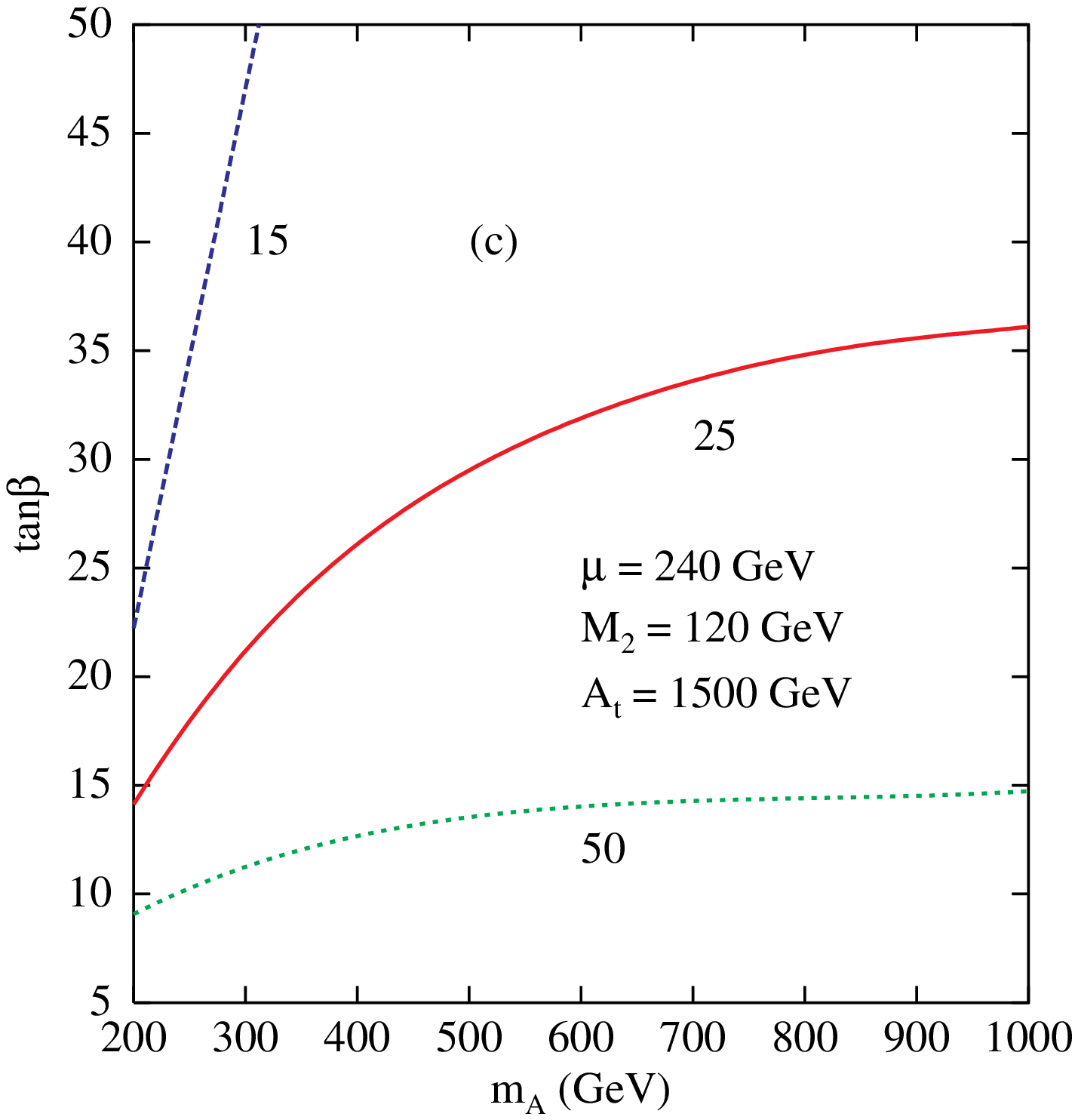}
\epsfxsize=6.5cm\epsfysize=6.0cm
                     \epsfbox{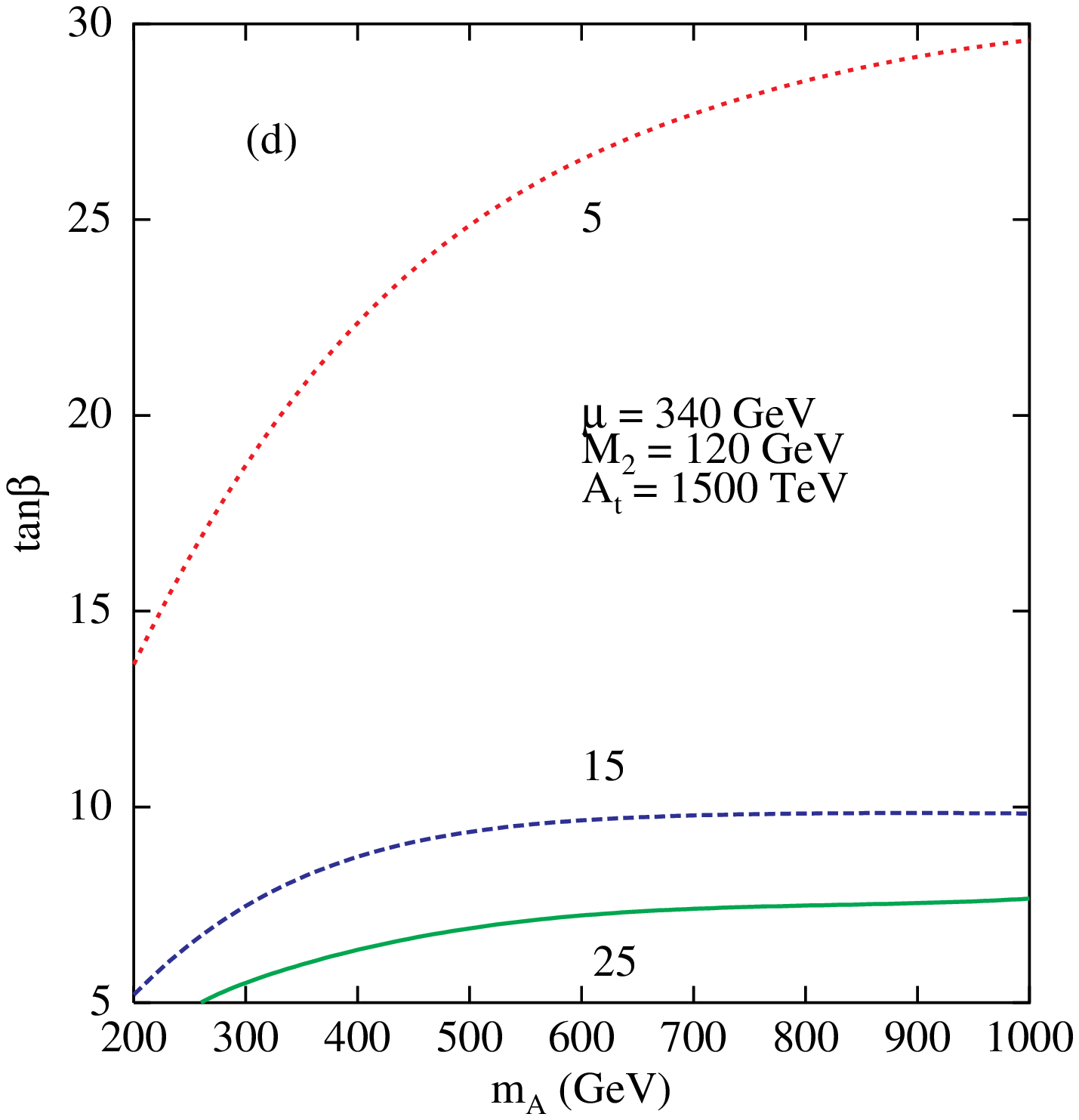}
}
\caption{{\it Contours of 2$j +$ 4$l$ events from $\neu$ pair
decay  via $\lambda$-coupling in the $m_A - \tan \beta$  plane.
Four panels are for different choices of $\mu$ and $A_t$.
}}
\label{fig:evts_ma_tanb_lambda}
\end{figure}

The $2j~+~4l~+~\not{E_T}$ events are subjected to the following event 
selection criteria \cite{rain_thesis}:

\begin{itemize}
\item {\bf Rapidity of jets:} for each jet, $2.5 \le |{\eta}| \le 5.0$. The 
upper limit is there to ensure detectability in the hadronic calorimeter. 
Also, the existence of two jets in opposite hemispheres is ensured by demanding
that $\eta_{j_1}\eta_{j_2}$ be negative.

\item {\bf Rapidity gap and isolation between jets:} $\Delta \eta_{{j_1}{j_2}}
 \ge 3.0$, $\Delta R_{{j_1}{j_2}} \ge 0.7$, 
where \\ $\Delta R^2 = \Delta \eta^2
 + \Delta \phi^2$, $\Delta \phi$ being the angular separation in the 
azimuthal plane. This retains  a large part of the signal, since the
signal rapidity interval tends to peak between 4 and 5.

\item {\bf Jet transverse energy:} for each forward jet, $E_T \ge 20~GeV$.

\item {\bf Central jet veto:} no jet in the large rapidity interval. 

\item {\bf Absence of b-jets:} no b-induced jet identified. Does not 
affect the signal at all.

\item {\bf Jet invariant mass:} $M_{{j_1}{j_2}} \ge 650~GeV$. Kills 
an enormous  amount of QCD background, and, as we shall see, helps in 
distinguishing the signal from other SUSY cascades.

\item {\bf Lepton location:} all four leptons to lie in the rapidity interval
between the forward-tagged jets, with $\Delta R_{lj} \ge 0.4$. 

\item {\bf Lepton transverse momentum(energy):} For each muon(electron), 
$p_T (E_T) \ge 20~ GeV$. Together with the isolation cut on leptons, 
this hardness cuts should eliminate of all SM backgrounds. 

\item {\bf Missing transverse energy}: ${\not{E}_T} \ge 15~GeV$ is required.
\end{itemize}

\begin{figure}[ht]
\centerline{
\epsfxsize=6.5cm\epsfysize=6.0cm
                     \epsfbox{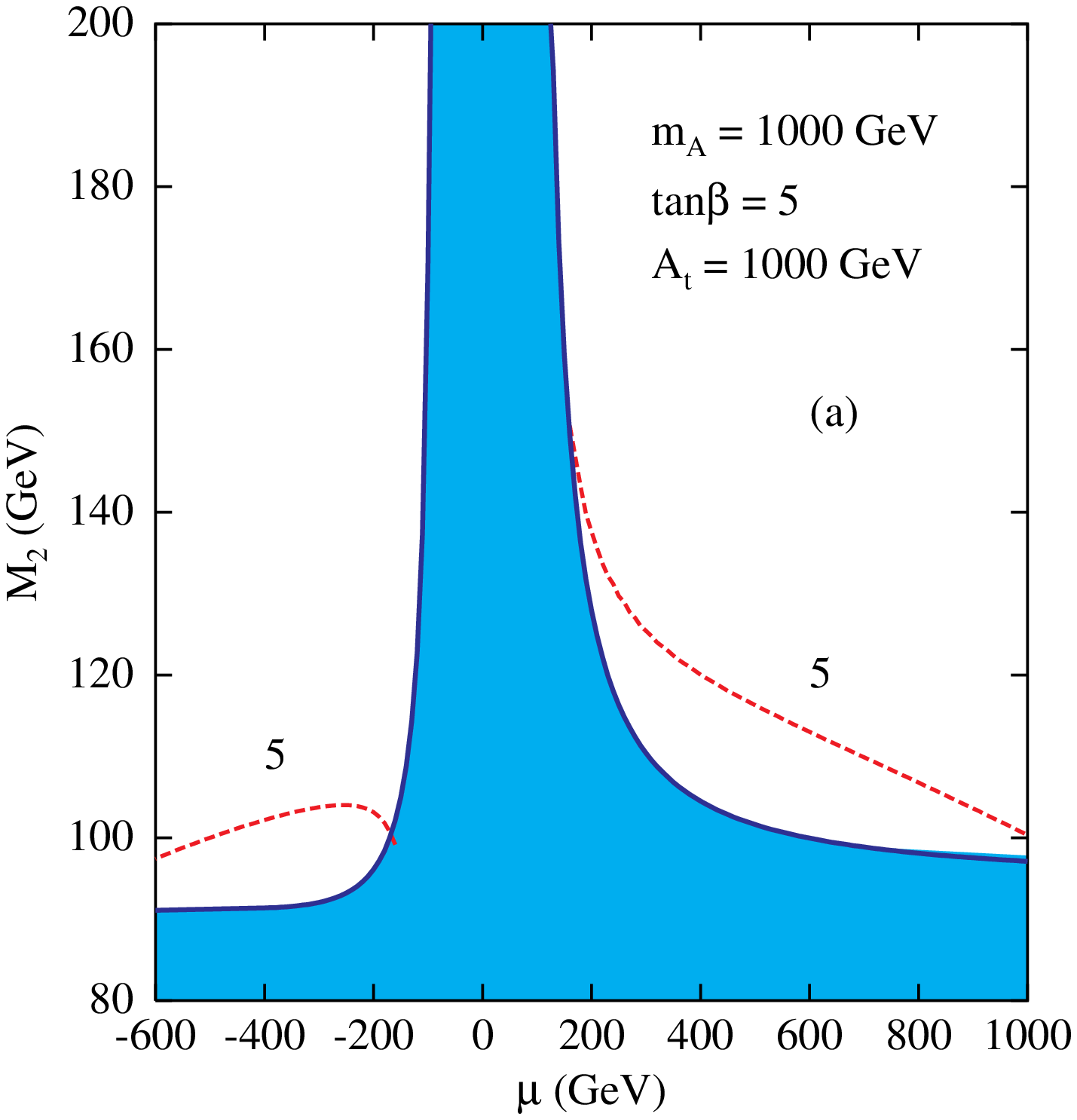}
        \hspace*{-1ex}
\epsfxsize=6.5cm\epsfysize=6.0cm
                     \epsfbox{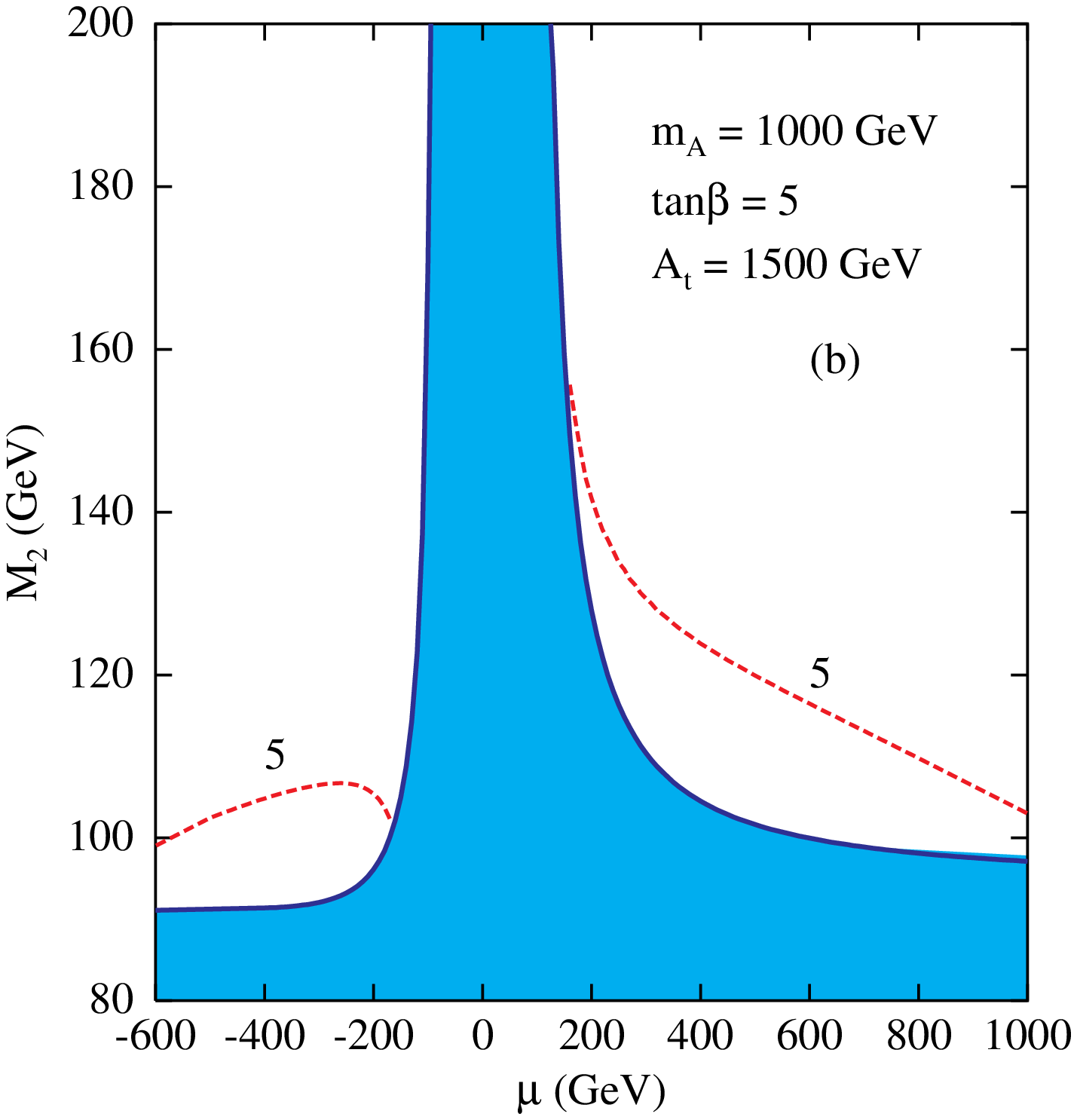}
}
\caption{{\it Contours of 5 events (2$j +$ 4$l$) in the $\mu - M_2$ plane coming 
  from $\neu$-pair decay via $\lambda$-coupling for
  a) $A_t$ = 1 $TeV$ and b) $A_t$ = 1.5 $TeV$. The shaded region is
disallowed from chargino search at LEP. In the region bounded by the
5-event contour and LEP-bound contour number of events is greater than
5.}}
\label{fig:evts_mu_m2_lambda}
\end{figure}

In figures \ref{fig:evts_ma_tanb_lambda}(a-d) we show contours in the
$m_A - \tan \beta$ plane for different numbers of events, predicted
for an integrated luminosity of 100 $fb^{-1}$. As can be seen from the
figures, up to about 50 events are predicted over a rather large
region of the parameter space. If the lepton $p_T (E_T)$ cut is
further relaxed to 15 $GeV$ (something that is feasible at the LHC)
\cite{atlas_cms}, the event rates are even higher. As expected from
the discussion in the previous section, the event rate tends to
decrease with a rise in $\tan \beta$, and increase with $m_A$. Also,
higher values of the soft breaking parameter $A_t$ seems to favour
high event rates. Side by side, a look at figures
\ref{fig:evts_mu_m2_lambda}(a-b) tells us how much of the LEP-allowed
region in the chargino-neutralino parameter space can be explored
through this channel. Judging by the fact that the signals are
practically free from standard model backgrounds, event rates of such
magnitude should be detectable. It is also interesting to note, by
comparison with figure 1, that even a small branching ratio for $h
\longrightarrow \chi_1^0 \chi_1^0$ can lead to useful signals at the
LHC.

\begin{figure}[ht]
\centerline{
\epsfxsize=6.5cm\epsfysize=6.0cm
                     \epsfbox{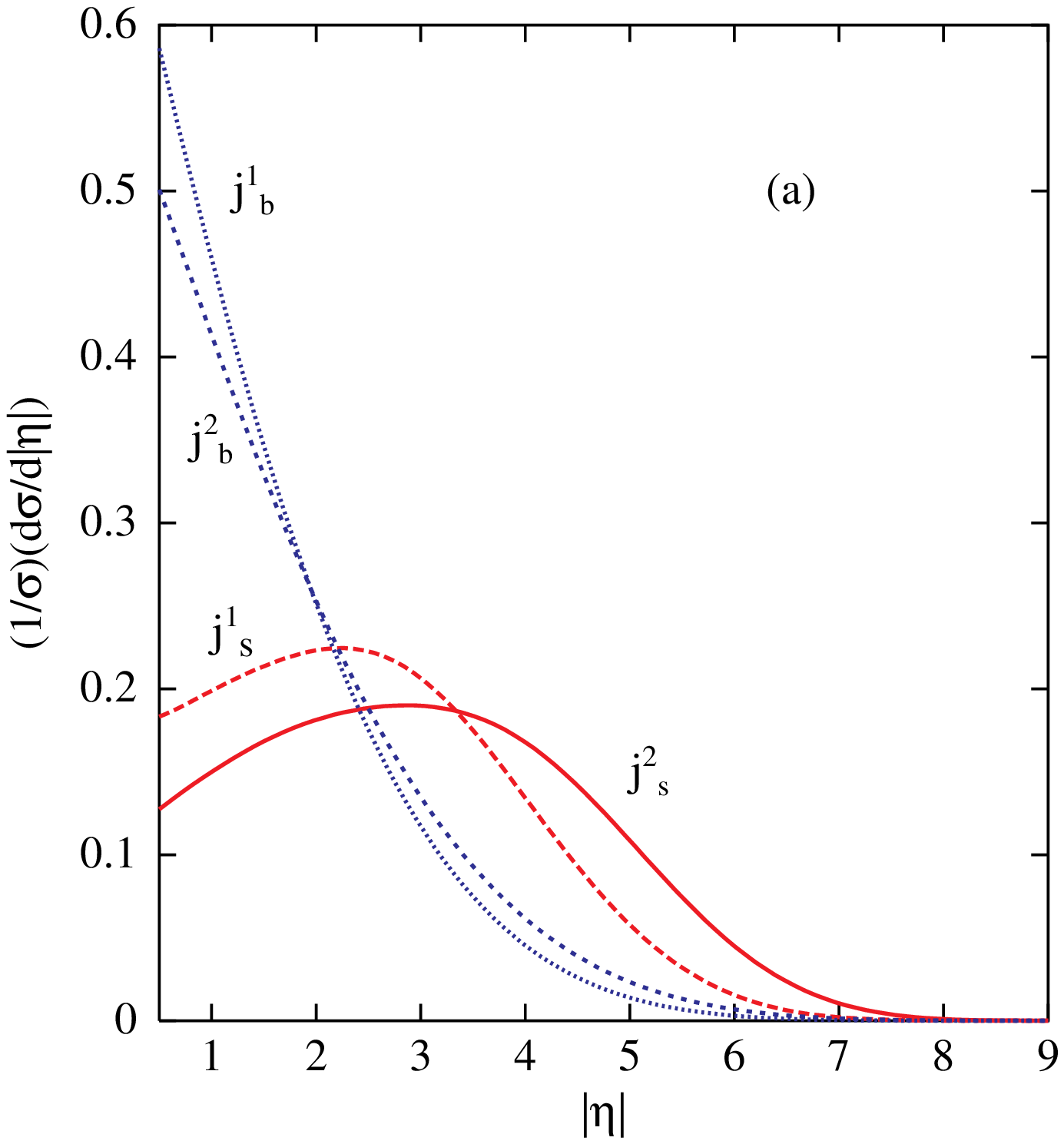}
\epsfxsize=6.5cm\epsfysize=6.0cm
                     \epsfbox{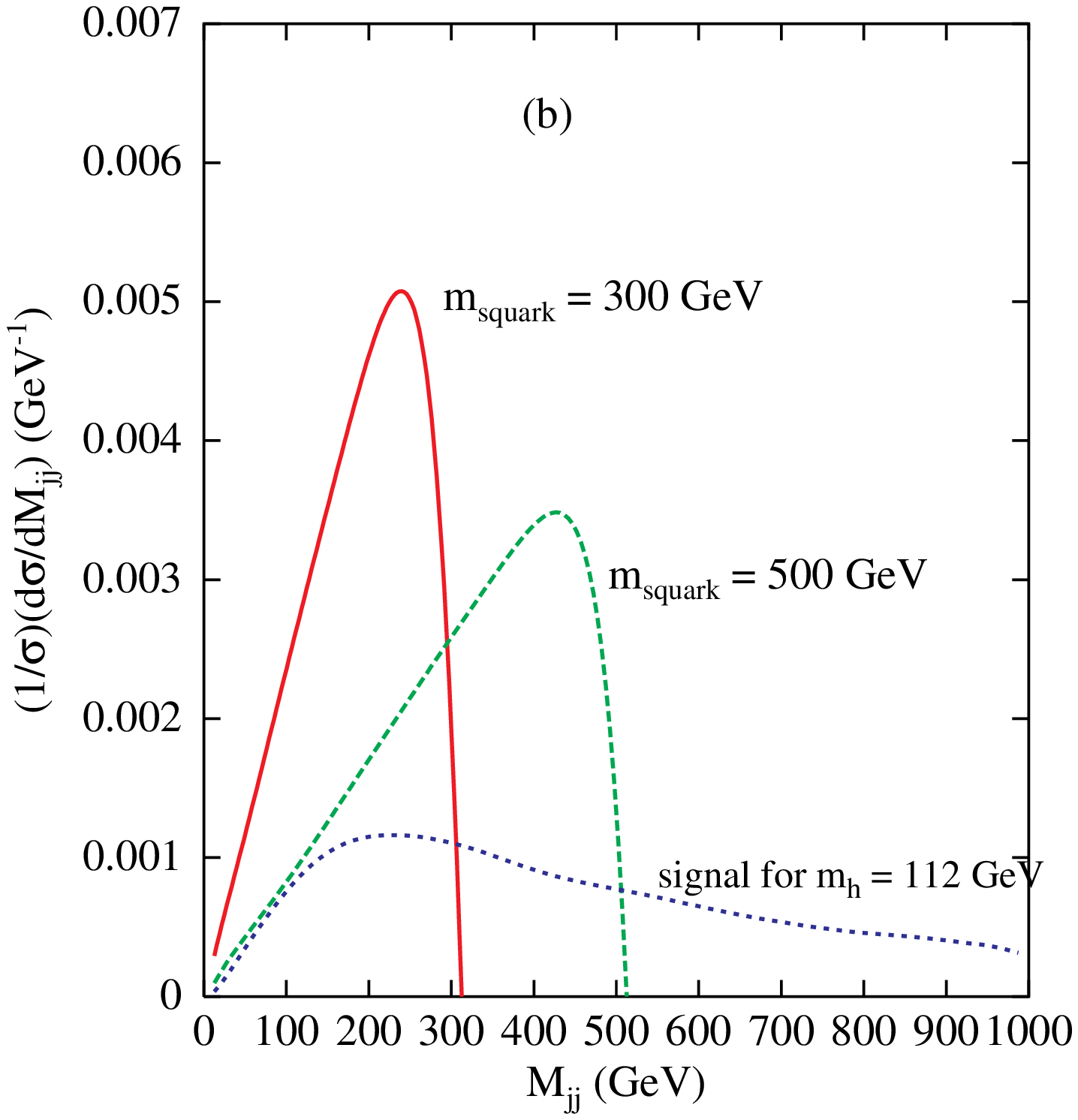}
}
\caption{{\it a) Normalised rapidity distributions of the jets form signal ($j^1_s , 
j^2_s$) and ``background'' ($j^1_b , j^2_b$) due to squark pair production and 
decay.
 b) Normalised invariant mass (of two jets) distribution of the 
  signal and ``background'' from squark production. Two different
  values of squark masses are chosen for the purpose of illustration.
 }}
\label{mas_rap}
\end{figure}

Finally, let us verify that the event selection procedure adopted here
enables us to differentiate the Higgs signals from remnants of
superparticle cascades. We demonstrate this by considering
$\chi_1^0$'s, together with quark jets, coming from pair-produced
squarks. Although the uncut cross-section for this process is
considerably larger than that for our signal, the forward-tagged jets
cause the latter to stand out.  In figure \ref{mas_rap}a we show the
rapidity distributions for the signal jets as well as for those
arising from squark decays. The latter exhibits a strong central
peaking, and is largely removed by a rapidity cut of 2.5. In addition,
if we look at the invariant mass distribution of the jet pair in
figure \ref{mas_rap}b, one can see that all such squark decay
`backgrounds' can be completely eliminated just with the invariant
mass cut of 650 $GeV$, while a substantial fraction of the signal
survives. This is true unless the squark mass is above 650 $GeV$. For
a squark mass of, say, 700 $GeV$, the uncut cross-section for
$\chi_1^0$-pair production through squark decay cascades is about 2-3
$pb$. However, as figure \ref{mas_rap}a tells us, the overwhelming
majority of the resulting jets lie in low-rapidity regions. Subjected
to the rapidity, $E_T$ and invariant mass cuts of the suggested
magnitudes, they get reduced to a level well below the threshold of
detectability. The same argument applies to cascade decays of gluino
pairs, where there is usually a greater multiplicity of quark jets,
and the probability of their merger into just two jets, satisfying all
the cuts, is extremely low.
\begin{center}
\begin{tabular}{|c|c|c|c|} 
\hline
$\mu$& $M_2$ &\multicolumn{2}{c|} {Number of} \\ 
(GeV)& (GeV)&\multicolumn{2}{c|} {Events} \\ 
\cline{3-4}
{}&{}&MSSM&$R_p\!\!\!\!\!\!/$ \\ \hline
200&137.6&0.037&0.55 \\ \hline
400 &120.02&0.039&0.44   \\ \hline
600&113.03&0.066&0.45 \\ \hline
800 &106.79&0.081&0.36  \\ \hline 
1000 &100.34&0.084&0.6  \\ \hline
-200&103.5&0.003&0.32   \\ \hline
-400 &102.18&0.101&0.3  \\ \hline 
-600&97.44&0.142&0.35 \\ \hline
\end{tabular}   
\end{center}
\noindent
{\bf Table 1} {\it The number of `$4l$ + forward jets' events coming from
neutralino pairs both for MSSM and with R-parity violation, for
different values of $\mu$ and $M_2$, with $\tan \beta$ = 5, $m_{\tilde q}
$ = 300 $GeV$ and $m_{\tilde l}$ = 200 $GeV$.}  

Similar final states can also in principle be faked by direct production
of neutralino pairs via VBF. There are two possibilities: ({\it a}) in 
an MSSM scenario, the pair-production of the second lightest neutralino, 
and their subsequent decays in the channel 
$\chi_2^0 \longrightarrow  \chi_1^0 l \overline{l}$ can lead to signals of
our type; and ({\it b}) the production of $\chi_1^0 \chi_1^0$,
$\chi_1^0 \chi_2^0$ or $\chi_2^0 \chi_2^0$ and their subsequent decays
(through $\lambda$-type couplings) including the possibility of 
R-parity violation can be the sources of similar signals. In table 1 we 
have shown the results of our estimate of such `fake' signals, obtained
at such points of the parameter space where the Higgs signals are at their
weakest (i.e. at about 5 events level). In the R-parity violating case,
one  $\lambda$-type coupling has been assumed, and has been kept at the
highest value compatible with phenomenological bounds. The forward jets
and leptons in between them have been subjected to the same cuts as the
ones employed for our Higgs signals. The table shows that after all cuts,
the number of such `backgrounds' get extremely suppressed, so that even
as low 5 as events for the Higgs signal should be quite conspicuous
compared to them.

\section{Signals with $\lambda'$-type interactions}

In presence of the $\lambda'$-type interactions (again, taken in
isolation), the lightest neutralino decays in the channel $\chi_1^0
\longrightarrow q \bar{q'} l$ or $\chi_1^0 \longrightarrow q \bar{q}
\nu_l$.  Of these, we use only the former channels where the decay
products are all visible. The signal will then consist of two forward
jets together with four central jets and two leptons, all in the
rapidity interval between the former. Obviously, the central jet veto
is not going to be effective here. However, a compensating feature
here is the visibility of all the particles in the final state.  Thus
two bunches of particles, each consisting of two jets and one lepton,
can be identified with the same invariant mass (equal to
$m_{\chi_1^0}$), and the whole bunch of particles in between the
forward-tagged jets can be reconstructed to an invariant mass peak
equal to the mass of the lighter neutral scalar $h$.
\begin{figure}[ht]
\centerline{
\epsfxsize=6.5cm\epsfysize=6.0cm
                     \epsfbox{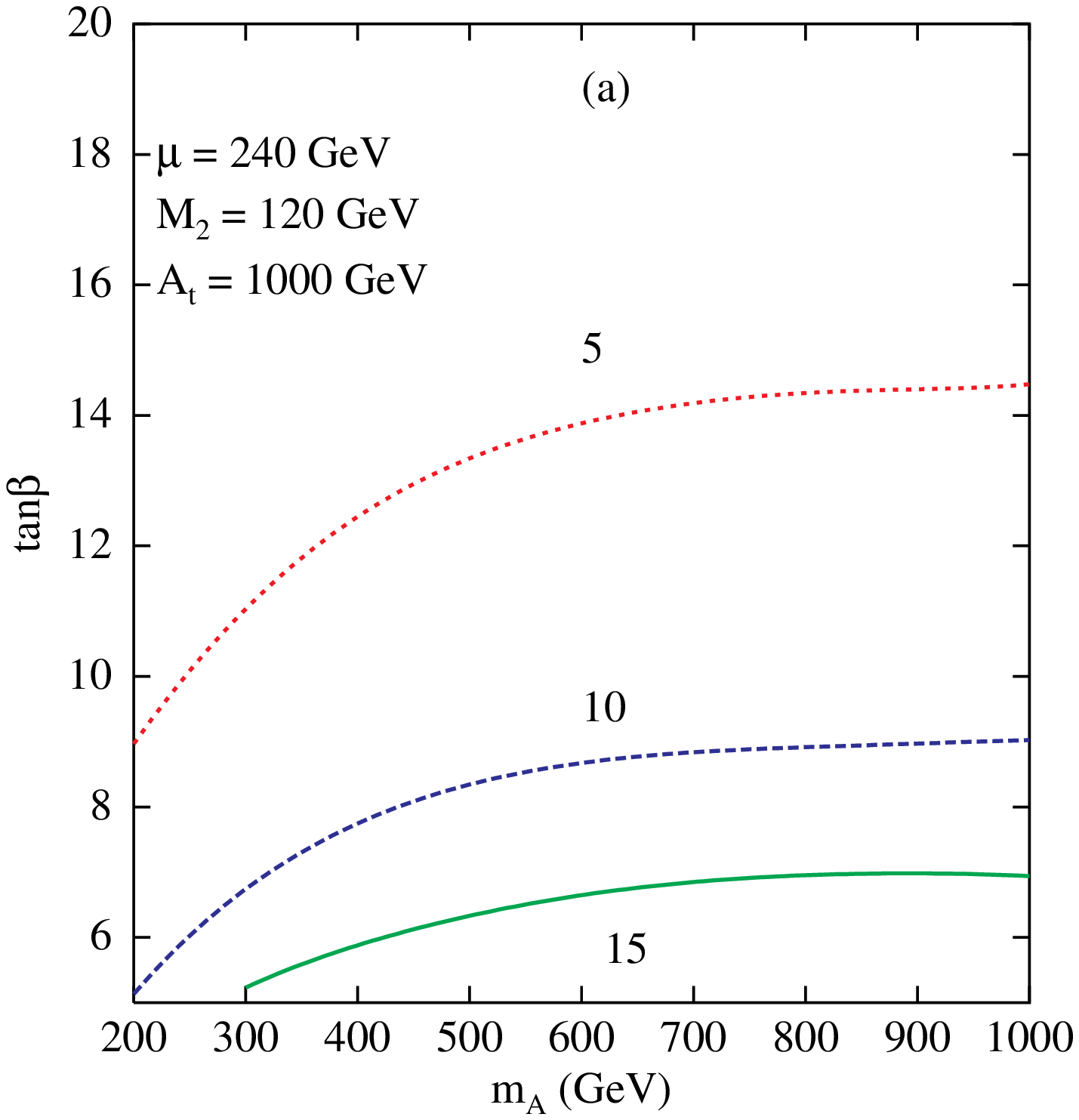}
        \hspace*{-1ex}
\epsfxsize=6.5cm\epsfysize=6.0cm
                     \epsfbox{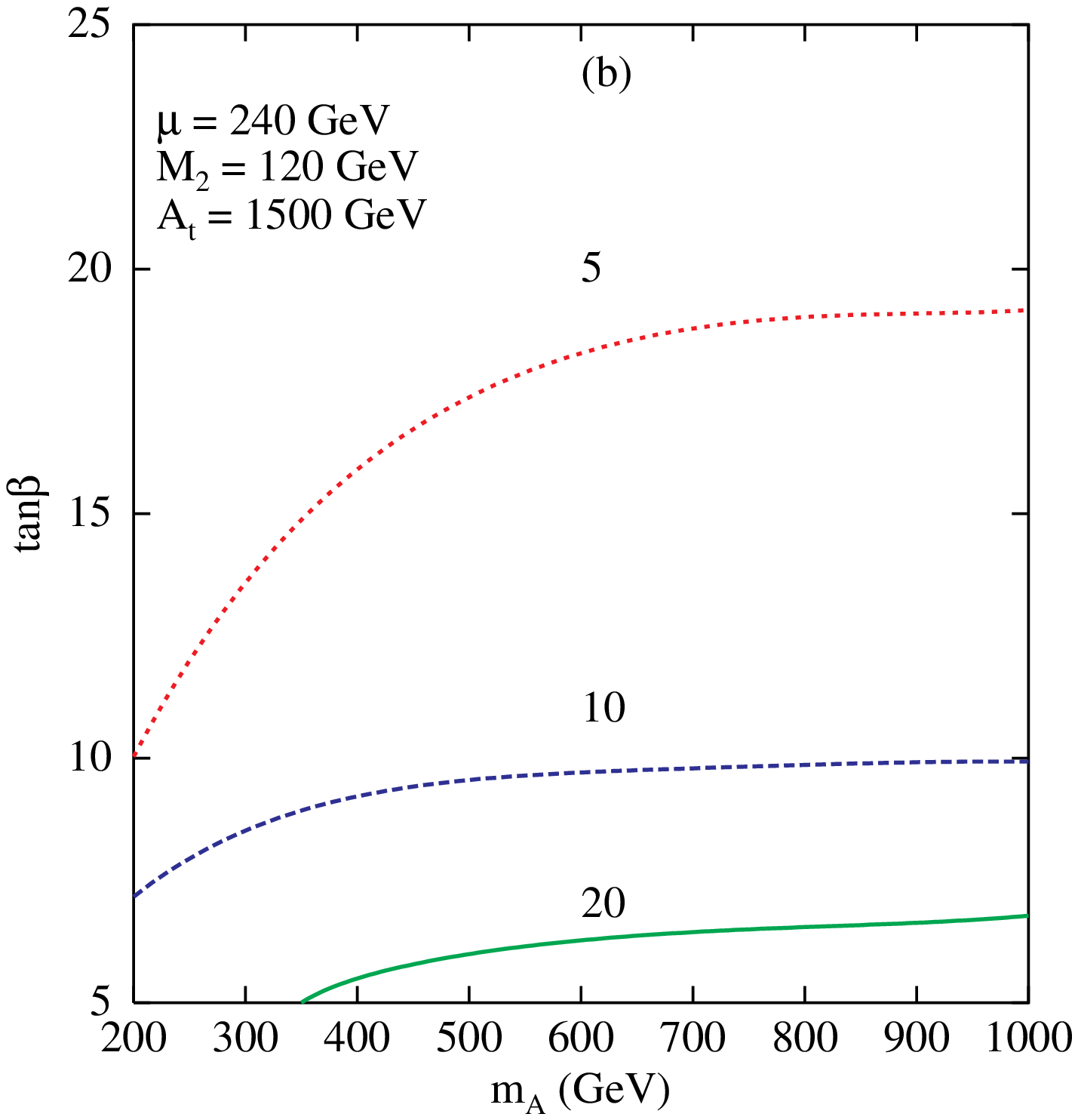}
}

\caption{{\it Contours  of $(jj)_f +$ central jets $+$ 2$l$ events (from $\neu$ pair
  decay via $\lambda '$-coupling) in the $m_A - \tan
  \beta$ plane, for two different choices of 
  $A_t$. }  }
\label{fig:ma_tanb_lp}
\end{figure}

The signal here thus corresponds to the process

\begin{equation}
qq \longrightarrow qqh \longrightarrow qq \chi_1^0 \chi_1^0 
\longrightarrow qq~~+~(4q)~+~2l
\end{equation}

\noindent 
leading to two forward-tagged jets with two, three or four jets (due to 
possible jet merger) together with two leptons in the rapidity
interval between the former. 

\begin{figure}[ht]
\centerline{
\epsfxsize=6.5cm\epsfysize=6.0cm
                     \epsfbox{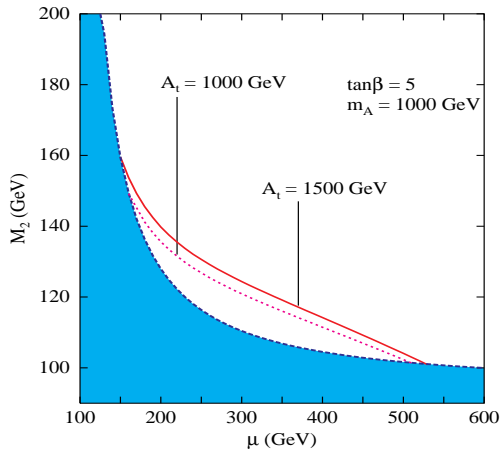}
}
\caption{{\it Contours of 5 events ($(jj)_f +$ central jets $+$ 2$l$) in 
the $\mu - M_2$ plane coming 
    from $\neu$-pair decay via $\lambda '$-coupling
    for $A_t$ = 1 $TeV$ and $A_t$ = 1.5 $TeV$. The shaded region is
    disallowed from chargino search at LEP. In the region bounded by
    the 5-event contour and LEP-bound contour, number of events is
    greater than 5.
}}  
\label{fig:mu_m2_lp}
\end{figure}

\begin{figure}[ht]
\centerline{
\epsfxsize=6.5cm\epsfysize=6.0cm
                     \epsfbox{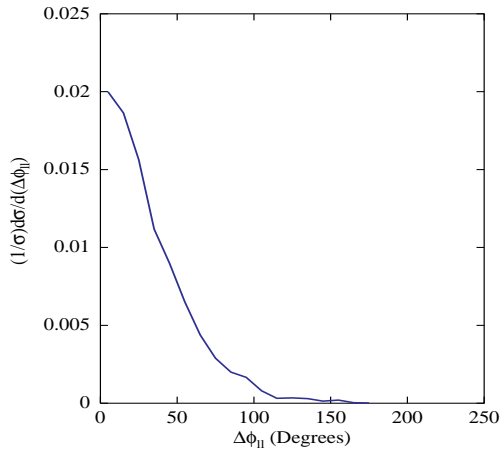}
}

\caption{{\it Normalised $\Delta \phi$ (angle between the leptons in the transverse
 plane) distribution of jets $+$ 2$l$ events coming from $\neu$ pair decay.}
}
\label{fig:deltaphi}
\end{figure}

In figures \ref{fig:ma_tanb_lp} and \ref{fig:mu_m2_lp} we present some
event contours, of the same style as those applied earlier. The two
forward jets are subjected to the same cuts as those applied earlier.
The rapidity interval demanded of the leptons and the central jets are
also the same. The jets arising from neutralino decays are also
required to have a minimum isolation of $\Delta R~=~0.6$ with respect
to the forward jets, and a minimum transverse energy of 20 $GeV$. The
minimum $p_{T}(E_{T})$ required of each lepton is 15 $GeV$.  In
addition, the invariant mass of the lepton pair is made to lie outside
an interval of $\pm 10~GeV$ of the $Z$-boson mass.  Also, we demand
that there should be no missing energy ($\le 10~GeV$), something that
enables us to distinguish the events from cases of undetected jets.
These cuts are found to be sufficient to eliminate standard model
background, including arising from Drell-Yan process and $t\bar{t}$
production.  Moreover, as figure \ref{fig:deltaphi} indicates, the
distribution in the azimuthal angle between the two leptons in the
transverse plane peaks at a very low value. Since dileptons from
Drell-Yan process tend to be aligned back-to-back, one kills all
backgrounds without affecting the signal strength by requiring this
angle to be less than $120$ degrees.

Clearly, we have fewer events predicted in this case than in
the one with the $\lambda$-type interactions. This is primarily due to 
the stringent cuts on the jets, in terms of both hardness and isolation
from the forward-tagged ones. Nonetheless, contours of 5 to 20 events span 
a substantial part of the parameter space. Considering the absence of SM 
backgrounds, these events should enable us to identify the Higgs in the
corresponding regions. The observed dependence on other parameters 
such as $A_t$ and $\mu$ is similar to that noted in the previous section.

\begin{figure}[ht]
\centerline{
\epsfxsize=6.5cm\epsfysize=6.0cm
                     \epsfbox{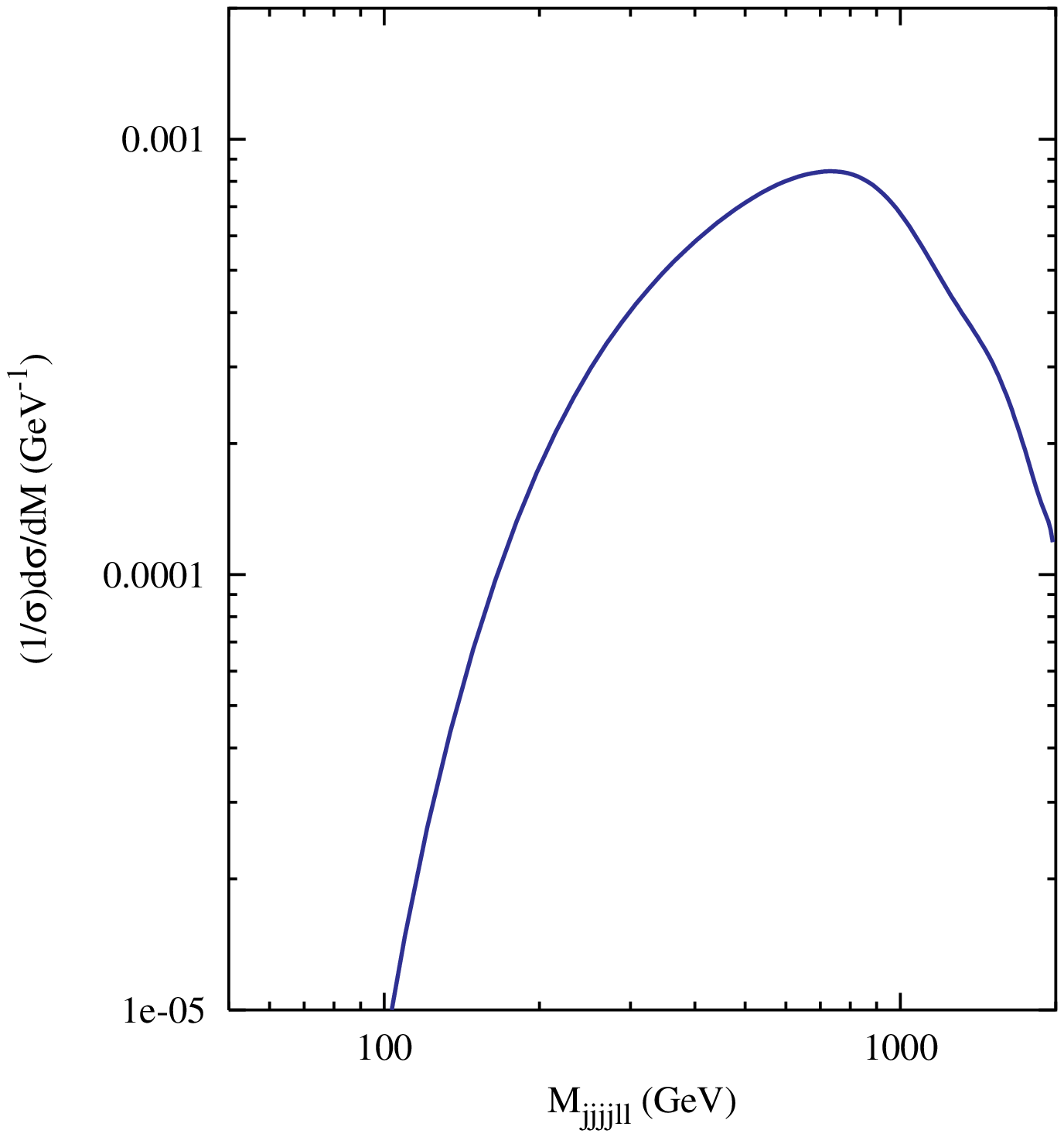}
}

\caption{{\it Normalised invariant mass   
 distribution of the visible products in between the forward tagged jets, 
from the decays of charginos and neutralinos, for $\mu$ = 500 GeV, $M_2$ = 100 GeV,
$\tan \beta$ = 5, $m_{\tilde q}$ = 300 $GeV$ and $m_{\tilde l}$ = 200 $GeV$.}  
}
\label{fig:invmas}
\end{figure}

Again, it is necessary to address the question of possible faking of
the signal through charginos and neutralinos. It is found that the
strongest candidates in this are $\chi_1^{+} \chi_1^{-}$ and
$\chi_2^{0} \chi_2^{0}$, when their subsequent R-parity violating
decays through $\lambda'$-type interactions are allowed. On the face
of it, such decays can have large branching ratios (of the order of
30-40\%) and the {\it forward jets + 2l + jets} events can have high
survival probability on imposing the aforementioned cuts. However, it
should be borne in mind that we are discussing a Higgs boson of mass
well below 135 $GeV$ or so. Thus the invariant mass of the visible
particles in between the forward-tagged jets should peak at that
value. On the other hand, simple kinematics tells us that the combined
invariant mass of the products of $\chi_1^{\pm}$ or $\chi_2^{0}$-pair
decays peaks at much higher values for  chargino and
neutralino masses currently allowed by the LEP data.  As an example,
we show in figure \ref{fig:invmas} the invariant mass distribution for
a case where the chargino mass is at the lowest limit.
In spite of that, the number of surviving events drop from about 85 to
0.01 once an invariant mass less than 140 $GeV$ is demanded from the
decay products. Thus such backgrounds, too, do not pose any hindrance
to unmasking the Higgs via the $\lambda'$-type interactions in the VBF
channel.

\begin{figure}[ht]
\centerline{
\epsfxsize=6.5cm\epsfysize=6.0cm
                     \epsfbox{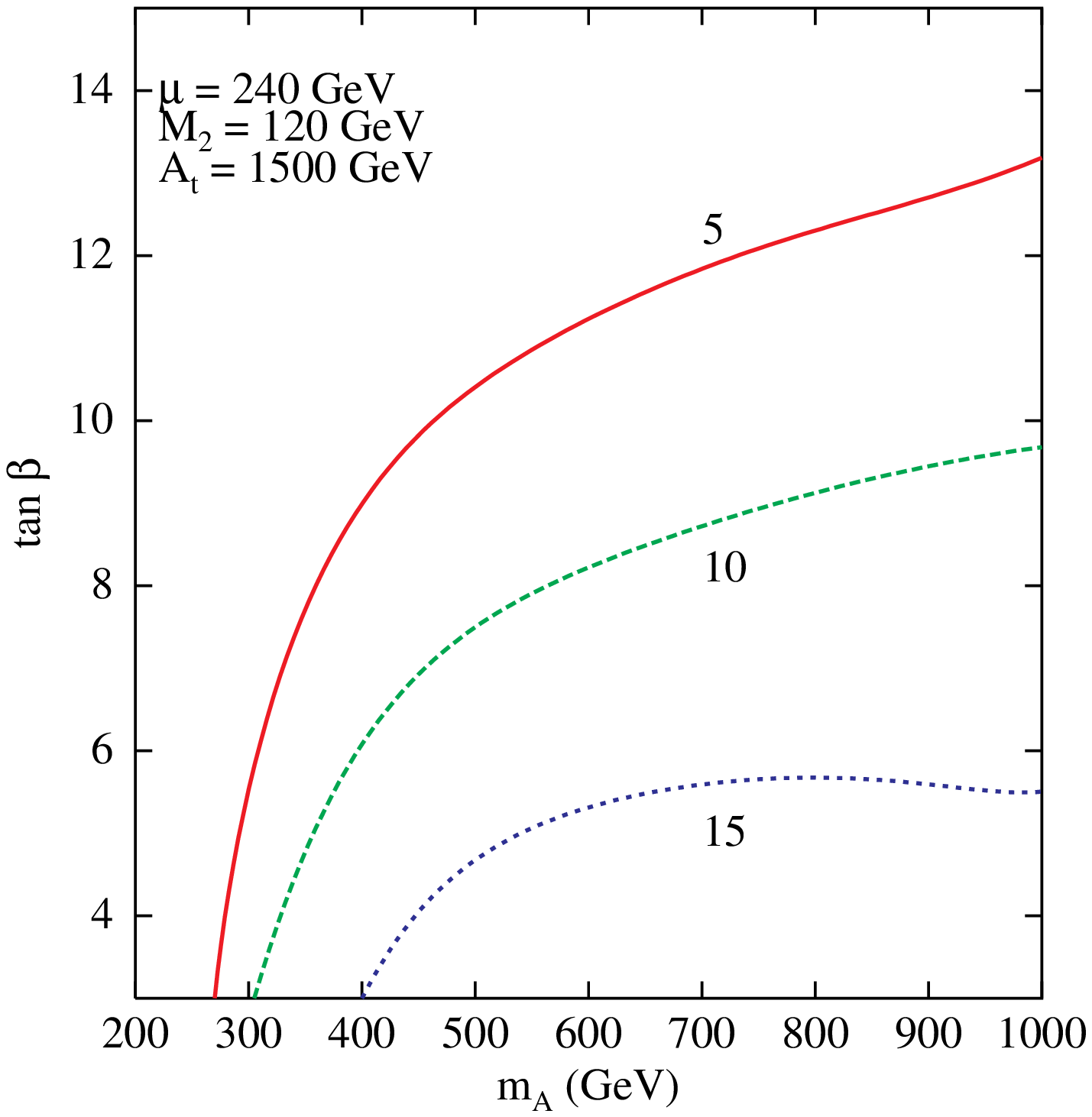}
}

\caption{\it {Contours of $(jj)_f +$ central jets $+$ 2$l$ events (from $\neu$ pair
  decay via $\lambda '$-coupling) in the $m_A - \tan
  \beta$ plane. Here we demand that the jets are well separated in the
 central region and the $p_T$-cut on the leptons is reduced to 10 $GeV$.}
}
\label{fig:ma_tanb_lp_pt10}
\end{figure}

Clearly, it requires considerable care to
reconstruct the neutralinos individually. In order to do that, one has
to select only those events where none of the two jets from one
neutralino merges with any jet arising from the decay of the other.
Using a jet merger criterion of $\Delta R \le 0.6$, we find that the
event rates get drastically reduced. One way to salvage them is to
relax the $p_{T}(E_{T})$ cut on the leptons. In figure
\ref{fig:ma_tanb_lp_pt10} we show how the rates are enhanced when a
separation of jets is demanded and for leptons the minimum
$p_{T}(E_{T})$ required is 10 $GeV$, with all other parameters
affecting the results in the same way as before.  Here one has two,
three or four-jet and dilepton events plus the forward-tagged jets. In
addition, the invariant mass of one lepton with one or two jets should
equal that of the remaining particles in the central region. This kind
of intertwining of leptons and jets in the invariant mass peaks makes
the signals almost completely free of SM backgrounds. In this way, the
lightest neutralino is also fully reconstructed. Since one can aspire
to see other signatures of such a neutralino at the LHC as well, the
reconstruction from the Higgs decay events with the neutralino mass
peaks at the right place serves in a big way to establish the {\it
  locus standi} of the process under investigation here, to remove
combinatoric backgrounds, and to improve measurements of the
Higgs-neutralino coupling.  Therefore, if the detector sensitivity
permits one to use these somewhat relaxed cuts, then one has one of
the cleanest signals of an intermediate mass Higgs in an R-parity
violating scenario, at least in the identified regions of the
parameter space.

\section{Summary and conclusions}

We have considered signals of the lightest neutral Higgs at the LHC in
a SUSY scenario where R-parity is violated through lepton number in
trilinear interactions. Regions have been identified in the parameter
space where the Higgs, lying the intermediate mass range, has a
perceptible decay width into a pair of lightest neutralinos. Then we
have considered the production of such a Higgs by the vector boson
fusion mechanism, and looked at the decay products of the two lightest
neutralinos, noting that the forward-tagged jets and the associated
event selection strategies remove interference from cascades arising
out of strongly interacting superparticles.

We have presented an analysis based on only the lepton number
violating trilinear interactions. Inclusion of the bilinear terms
$\epsilon_i L_i H_2$ in the superpotential will open additional decay
channels of the lightest neutralino like $\chi_1^0 \longrightarrow l W$
and $\chi_1^0 \longrightarrow \nu Z$ through neutralino-neutrino and
charged lepton-chargino mixing \cite{blp}. However, since we are
concerned here with a parameter region where the decay of $\chi_1^0$
can lead only to virtual $W$ and $Z$, the final products will still
consist of three fermions. In this case, events of both the types
discussed in sections 3 and 4 will be always present. Another
consequence of the bilinear terms is mixing between the neutral Higgs
and sneutrinos in the scalar potential. Such mixing may somewhat alter
the parameter space discussed in section 2, but no qualitative
difference is expected, given the phenomenological constraints on
models of this type \cite{bi_rp_pheno}.  Therefore, the existence of
bilinear R-violating interactions are expected to result in signals of
the same kind as those investigated here.

We observe that while the signals obtained from the $\lambda$-type
interactions are more copious and cover a larger area of the parameter
space, the $\lambda'$-type interactions provide a way of
reconstructing the Higgs completely, though  they lead to smaller event 
rates after applying the cuts. The complete reconstruction of
the events including the lightest neutralinos, however, require cuts
that tend to reduce the event rates. It is by being able to identify
leptons with transverse energies down to $10~GeV$ that one can
reconstruct both the Higgs and the neutralinos which act as
intermediaries in the signals of our interest. Therefore, if optimal
detection efficiencies of the various final states discussed here can
be achieved at the LHC, it will be of great help in identifying an
intermediate mass Higgs in an R-parity violating supersymmetric
theory. In addition, by looking for signals of this kind, we can
obtain useful information on the different couplings of an
intermediate Higgs boson when it is discovered, so as to be
enlightened on what kind of an electroweak symmetry breaking scenario
it represents.

{\bf Acknowledgements}: B.M. thanks Dieter Zeppenfeld for useful
comments, and acknowledges the hospitality of the Theory Division,
CERN, where part of this work was done. Both the authors gratefully
acknowledge the help received from Debajyoti Choudhury in preparing
the figures.

\newcommand{\plb}[3]{{Phys. Lett.} {\bf B#1} #2 (#3)}                  %
\newcommand{\prl}[3]{Phys. Rev. Lett. {\bf #1} #2 (#3)}        %
\newcommand{\rmp}[3]{Rev. Mod.  Phys. {\bf #1} #2 (#3)}             %
\newcommand{\prep}[3]{Phys. Rep. {\bf #1} #2 (#3)}                     %
\newcommand{\rpp}[3]{Rep. Prog. Phys. {\bf #1} #2 (#3)}             %
\newcommand{\prd}[3]{{Phys. Rev.}{\bf D#1} #2 (#3)}                    %
\newcommand{\np}[3]{Nucl. Phys. {\bf B#1} #2 (#3)}                     %
\newcommand{\npbps}[3]{Nucl. Phys. B (Proc. Suppl.) 
           {\bf #1} (#3) #2}                                           %
\newcommand{\sci}[3]{Science {\bf #1} #2 (#3)}                 %
\newcommand{\zp}[3]{Z.~Phys. C{\bf#1} #2 (#3)}                 %
\newcommand{\mpla}[3]{Mod. Phys. Lett. {\bf A#1} #2 (#3)}             %

\newcommand{\astropp}[3]{Astropart. Phys. {\bf #1} #2 (#3)}            %
\newcommand{\ib}[3]{{\em ibid.\/} {\bf #1} #2 (#3)}                    %
\newcommand{\nat}[3]{Nature (London) {\bf #1} (#3) #2}         %
\newcommand{\nuovocim}[3]{Nuovo Cim. {\bf #1} #2 (#3)}         %
\newcommand{\yadfiz}[4]{Yad. Fiz. {\bf #1} (#3) #2 [English            %
        transl.: Sov. J. Nucl.  Phys. {\bf #1} #3 (#4)]}               %
\newcommand{\philt}[3]{Phil. Trans. Roy. Soc. London A {\bf #1} #2  
        (#3)}                                                          %
\newcommand{\hepph}[1]{(hep--ph/#1)}           %
\newcommand{\hepex}[1]{(electronic archive:     hep--ex/#1)}           %
\newcommand{\astro}[1]{(electronic archive:     astro--ph/#1)}         %
\newcommand{\epj}[3]{Euro. Phys. J {\bf C#1} #2 (#3)}  



\begin{thebibliography}{99}
\bibitem{atlas_cms} Lepton ($e$ or $\mu$) can be identified at the
LHC even with a $p_T$ ($E_T$) cut of 10 $GeV$. For example see \cite{smear}.
\end{thebibliography}
\end{document}